\documentclass[prb,showpacs,twocolumn,floats,10pt,aps,citeautoscript,longbibliography,superscriptaddress]{revtex4-2}
\usepackage{dcolumn}
\usepackage{anyfontsize} 
\usepackage{microtype} 
\usepackage{graphicx} 
\usepackage{xspace} 

\usepackage[usenames,dvipsnames]{xcolor} 

\usepackage{xparse} 

\usepackage{algorithm}
\usepackage{algpseudocode}

\usepackage[normalem]{ulem} 
\definecolor{darkgreen}{rgb}{0,0.5,0}
\definecolor{purple}{rgb}{0.6,0,0.5}
\definecolor{orange}{rgb}{1,0.5,0}
\definecolor{darkred}{rgb}{.7,0,0}
\definecolor{darkblue}{rgb}{0,0,.6}
\definecolor{grey}{rgb}{.6,.6,.6}
\definecolor{dimgreen}{rgb}{0.2,0.7,0.2}

\newcommand{\TriangleEllipseGrey}{\protect\raisebox{-0.5mm}{\protect\includegraphics[width=0.07\linewidth]{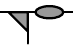}}}
\newcommand{\CircleWhiteC}{\protect\raisebox{-0.5mm}{\protect\includegraphics[width=0.037\linewidth]{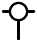}}}
\newcommand{\CircleGreyC}{\protect\raisebox{-0.5mm}{\protect\includegraphics[width=0.037\linewidth]{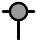}}}
\newcommand{\TriangleWhiteA}{\protect\raisebox{-0.5mm}{\protect\includegraphics[width=0.037\linewidth]{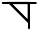}}}
\newcommand{\TriangleWhiteB}{\protect\raisebox{-0.5mm}{\protect\includegraphics[width=0.037\linewidth]{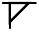}}}
\newcommand{\jvd}[1]{{\color{darkblue}{#1}}}

\newcommand{\jvdomit}[1]{}

\usepackage{amsmath} 
\usepackage{amssymb}
\usepackage{mathrsfs} 
\usepackage{bbm}  
\renewcommand{\vec}[1]{{\boldsymbol{#1}}} 
\newcommand{\mr}[1]{\ensuremath{\mathrm{#1}}}
\newcommand{\mc}[1]{\ensuremath{\mathcal{#1}}}
\allowdisplaybreaks 
\usepackage{enumitem} 

\usepackage{scalefnt} 

\usepackage{tikz}
\usepackage{algorithm}
\usepackage{algpseudocode}

\usepackage[colorlinks, citecolor={blue!50!black}, urlcolor={blue!50!black}, linkcolor={red!50!black}]{hyperref}

\usepackage{scalerel,stackengine}  

\newcommand{\Eq}[1]{Eq.~\eqref{#1}}
\newcommand{\Eqs}[1]{Eqs.~\eqref{#1}}

\def\Pc{\mathcal{P}}
\def\Qc{\mathcal{Q}}
\def\Hc{\mathcal{H}}

\def\Pzero{\mathcal{P}^{0\perp}}
\def\Pone{\mathcal{P}^{1\perp}}
\def\Ptwo{\mathcal{P}^{2\perp}}
\def\Pbond{\mathcal{P}^\mbond}
\def\Pzerosite{\mathcal{P}^\mzerosite}
\def\Ponesite{\mathcal{P}^\monesite}
\def\Ptwosite{\mathcal{P}^\mtwosite}

\def\Pnsite{\mathcal{P}^{n\mathrm{s}}}
\def\Pnpsite{\mathcal{P}^{n'\mathrm{s}}}
\def\Pnminusonesite{\mathcal{P}^{(n-1)\mathrm{s}}}
\def\Pn{\mathcal{P}^{n\perp}}
\def\Pnp{\mathcal{P}^{n'\perp}}
\def\Pk{\mathcal{P}^\ks}
\def\Pd{\mathcal{P}^\ds}
\def\Px{\mathcal{P}^\xs}
\def\Pxb{\mathcal{P}^{\xsb}}
\def\Qk{\mathcal{Q}^\ks}
\def\Qd{\mathcal{Q}^\ds}
\def\Qx{\mathcal{Q}^\xs}
\def\Qxb{\mathcal{Q}^{\xsb}}
\def\Pkk{\mathcal{P}^{\ks\ks}}
\def\Pkd{\mathcal{P}^{\ks\ds}}
\def\Pdk{\mathcal{P}^{\ds\ks}}
\def\Pdd{\mathcal{P}^{\ds\ds}}
\def\Pxxb{\mathcal{P}^{\xs\xsb}}
\def\Pxe{\mathcal{P}^{\xs\es}}
\def\Pex{\mathcal{P}^{\es\xs}}
\def\Pke{\mathcal{P}^{\ks\es}}
\def\Pek{\mathcal{P}^{\es\ks}}
\def\Pde{\mathcal{P}^{\ds\es}}
\def\Ped{\mathcal{P}^{\es\ds}}
\def\doubleVzerosite{\doubleV^\mzerosite}
\def\doubleVonesite{\doubleV^\monesite}
\def\doubleVtwosite{\doubleV^\mtwosite}
\def\doubleVnsite{\doubleV^\mnsite}
\def\doubleVzero{\doubleV^{0\perp}}
\def\doubleVone{\doubleV^{1\perp}}
\def\doubleVtwo{\doubleV^{2\perp}}
\def\doubleVn{\doubleV^{n\perp}}
\def\Rc{\mc{R}}
\def\Lc{\mc{L}}

\newcommand{\ellminusone}{{\ell \hspace{-0.2mm}-\hspace{-0.2mm} 1}}
\newcommand{\ellplusone}{{\ell  \hspace{-0.2mm}+\hspace{-0.2mm} 1}}
\newcommand{\ellplustwo}{{\ell  \hspace{-0.2mm}+\hspace{-0.2mm} 2}}

\newcommand{\tell}{{\tilde \ell}}
\newcommand{\ellp}{{\ell'}}
\newcommand{\tellp}{{{\tilde \ell}'}}
\newcommand{\eLL}{{\mbox{\small$\mathscr{L}$}}}
\newcommand{\seLL}{{\scriptscriptstyle \! \mathscr{L}}}
\newcommand{\scripteLL}{{\scriptstyle \! \mathscr{L}}}
\newcommand{\sseLL}{{{\mbox{\tiny$\!\mathscr{L}$}}}}

\def\Lc{\mathcal{L}}

\def\Ab{{\overline{A}}}

\def\Bb{{\overline{B}}}

\def\Dast{D^\ast}

\def\Db{\overline{D}}

\def\Psik{\Psi^\ks} 
\def\Psid{\Psi^\ds} 
\def\Phik{\Phi^\ks} 
\def\Phid{\Phi^\ds}

\newcommand*{\ndots}{\kern-0.075em.\kern-0.05em.\kern-0.05em.}  
\newcommand*{\nidots}{.\kern-0.05em.\kern-0.05em.} 
\newcommand*{\ncdots}{\kern-0.15em\cdot\kern-0.2em\cdot\kern-0.2em\cdot\kern-0.15em}  

\newcommand{\bra}[1]{\ensuremath{\langle #1 |}}
\newcommand{\ket}[1]{\ensuremath{| #1 \rangle}}
\newcommand{\prj}[1]{\ensuremath{| #1 \rangle\!\langle #1 |}}
\newcommand{\ovl}[2]{\ensuremath{\langle #1 | #2 \rangle}}

\newcommand{\pdag}{{\protect\vphantom{dagger}}}

\def\ellb{{\bar{\ell}}}
\def\ellbp{{\bar{\ell}{}'}}

\newcommand{\im}{\mathrm{im}}

\newcommand{\zerosite}{\textrm{0s}}
\newcommand{\onesite}{\textrm{1s}}
\newcommand{\twosite}{\textrm{2s}}
\newcommand{\nsite}{{$n$\textrm{s}}}

\newcommand{\mxsite}{\mathrm{x}}
\newcommand{\mzerosite}{\mathrm{0s}}
\newcommand{\monesite}{\mathrm{1s}}
\newcommand{\mtwosite}{\mathrm{2s}}
\newcommand{\mnsite}{{n\textrm{s}}}
\newcommand{\mnminusonesite}{{(n-1)\textrm{s}}}
\newcommand{\mbond}{\mathrm{b}}

\NewDocumentCommand{\doubleI}{O{}}{\mathbbm{1}_{#1}}
\NewDocumentCommand{\doubleIb}{O{}}{{\overline{\mathbbm{1}}_{#1}}}
\NewDocumentCommand{\doubleIk}{O{}}{\mathbbm{1}^\ks_{\! #1}}
\NewDocumentCommand{\doubleId}{O{}}{\mathbbm{1}^\ds_{\! #1}}
\NewDocumentCommand{\doubleIp}{O{}}{\mathbbm{1}^\ps_{\! #1}}
\NewDocumentCommand{\doubleV}{O{}}{\mathbbm{V}_{\! #1}}
\NewDocumentCommand{\doubleVk}{O{}}{\mathbbm{V}^\ks_{\! #1}}
\NewDocumentCommand{\doubleVd}{O{}}{\mathbbm{V}^\ds_{\! #1}}
\NewDocumentCommand{\doubleVp}{O{}}{\mathbbm{V}^\ps_{\! #1}}
\NewDocumentCommand{\doublev}{o}{{\mathbbm{v}_{#1}}}
\NewDocumentCommand{\doubleVb}{o}{{\overline{\mathbbm{V}}_{\! #1}}}
\NewDocumentCommand{\doubleVt}{o}{{\widetilde{\mathbbm{V}}_{\! #1}}}
\NewDocumentCommand{\doubleVh}{o}{\widehat{{\mathbbm{V}}_{\! #1}}}
\NewDocumentCommand{\doubleW}{o}{\mathbbm{W}_{\! #1}}
\NewDocumentCommand{\doubleWk}{o}{\mathbbm{W}^\ks_{\! #1}}
\NewDocumentCommand{\doubleWd}{o}{\mathbbm{W}^\ds_{\! #1}}
\NewDocumentCommand{\doubleWb}{o}{{\overline{\mathbbm{W}}_{\! #1}}}
\NewDocumentCommand{\doubleWt}{o}{{\widetilde{\mathbbm{V}}_{\! #1}}}
\NewDocumentCommand{\doubleWh}{o}{{\widehat{\mathbbm{V}}_{\! #1}}}

\def\X{{\scriptstyle {\rm X}}}
\def\D{{\scriptstyle {\rm D}}} 
\def\DD{{\scriptstyle {\rm DD}}} 
\def\E{{\scriptstyle {\rm E}}} 
\def\K{{\scriptstyle {\rm K}}} 
\def\P{{\scriptstyle {\rm P}}} 
\def\xs{{\scriptscriptstyle {\rm X}}}
\def\xsb{{\overline {\scriptscriptstyle {\rm X}}}}
\def\xsp{{\scriptscriptstyle {\rm X}'}}
\def\xsbp{{\overline {\scriptscriptstyle {\rm X}}{}'}}

\def\ds{{\scriptscriptstyle {\rm D}}}
\def\ks{{\scriptscriptstyle {\rm K}}}

\def\es{{\scriptscriptstyle {\rm E}}}

\def\ps{{\scriptscriptstyle {\rm P}}}

\def\Xb{{\overline{\mathrm{\scriptstyle X}}}}

\def\variance{\Delta_{E}}

\NewDocumentCommand{\cor}{mod()}
{
	#1\IfValueTF{#2}{[#2]}{}\IfValueTF{#3}{(#3)}{}
}

\usepackage{mathtools}

\makeatletter
\def\maketitle{
\@author@finish
\title@column\titleblock@produce
\suppressfloats[t]}
\makeatother

\begin{document} 

\title{Projector formalism for kept and discarded spaces of matrix product states}

\author{Andreas Gleis}
\affiliation{Arnold Sommerfeld Center for Theoretical Physics, 
Center for NanoScience,\looseness=-1\,  and 
Munich Center for \\ Quantum Science and Technology,\looseness=-2\, 
Ludwig-Maximilians-Universit\"at M\"unchen, 80333 Munich, Germany}
\author{Jheng-Wei Li}
\affiliation{Arnold Sommerfeld Center for Theoretical Physics, 
Center for NanoScience,\looseness=-1\,  and 
Munich Center for \\ Quantum Science and Technology,\looseness=-2\, 
Ludwig-Maximilians-Universit\"at M\"unchen, 80333 Munich, Germany}
\author{Jan von Delft}
\affiliation{Arnold Sommerfeld Center for Theoretical Physics, 
Center for NanoScience,\looseness=-1\,  and 
Munich Center for \\ Quantum Science and Technology,\looseness=-2\, 
Ludwig-Maximilians-Universit\"at M\"unchen, 80333 Munich, Germany}

\begin{abstract}
\begin{center}
(Dated: \today)
\end{center}

Any matrix product state $\ket{\Psi}$ has a set of associated kept and discarded spaces, needed 
for the description of $\ket{\Psi}$, and changes thereof, respectively. These induce a partition of the
full Hilbert space of the system into mutually orthogonal spaces of irreducible $n$-site variations of $\ket{\Psi}$.
Here, we introduce a convenient projector formalism and diagrammatic notation to characterize these $n$-site spaces explicitly. This greatly facilitates the formulation of MPS algorithms that explicitly or implicitly employ discarded spaces. As an illustration, we derive an explicit expression for the $n$-site energy variance and evaluate it numerically for a model with long-range hopping. We also describe an efficient algorithm for computing low-lying $n$-site excitations above a finite MPS ground state.
\\
\\
\noindent
DOI:
\end{abstract}

\maketitle

\section{Introduction}
\label{sec:introduction}

Matrix product states (MPS) are widely used for the numerical description of quantum systems defined on one- or two-dimensional lattices. Well-known MPS-based algorithms include ground state searches and time evolution using the density matrix renormalization group (DMRG and tDMRG)
\cite{White1992,White1993,Daley2004,White2005,Hubig2015,Hubig2018}, time-evolving block decimation (TEBD) methods \cite{Vidal2003,Vidal2004,Vidal2007a}, or the 
time-dependent variational principle (TDVP) \cite{Haegeman2011,Lubich2015a,Haegeman2016,ZaunerStauber2018,Vanderstraeten2019}; and the computation of spectral information 
using the numerical renormalization group (NRG)  
\cite{Wilson1975,Peters2006,Weichselbaum2007}, DMRG \cite{Hallberg1995,Kuhner1999,Jeckelmann2002,Holzner2011},  or so-called post-MPS approaches \cite{Haegeman2013a,Vanderstraeten2019};
see Refs.~\cite{Schollwoeck2011,Weichselbaum2012a,Paeckel2019} for reviews. 

All such algorithms involve update steps: a quantum state of interest, $\ket{\Psi}$, is represented in MPS form, and its constituent tensors are updated, e.g.\  during optimization or time evolution. During an update, highly relevant information is \textit{kept} ($\K$) 
and less relevant information \textit{discarded} ($\D$). A sequence of updates thereby endows the full Hilbert space of the system, $\doubleV$, with a structure of intricately nested $\K$ or $\D$\ subspaces, changing with each update, containing states from $\doubleV$ which either do ($\K$) or do not ($\D$) contribute to the description of $\ket{\Psi}$. 

The nested structure of $\doubleV$ is rarely made explicit in the formulation of MPS algorithms. A notable exception is NRG, where $\D$ states are used to construct a complete basis \cite{Anders2005} of approximate energy eigenstates for $\doubleV$, facilitating the computation of time evolution or spectral information \cite{Peters2006,Weichselbaum2007}. For the computation of local multipoint correlators \cite{Lee2021} using NRG, it has proven useful to  elucidate the structure of $\K$ and $\D$ subspaces by introducing projectors having these subspaces as their images. The orthogonality properties of $\K$ and $\D$ projectors bring structure and clarity to the description of rather complex algorithmic strategies. 

Inspired by the convenience of $\K$ and $\D$ projectors in the context of NRG, we here introduce an analogous but more general $\K$,$\D$ projector formalism and diagrammatic conventions suitable for the description of arbitrary MPS algorithms. In particular, our $\K$,$\D$ projectors offer a natural language for the formulation of algorithms that explicitly or implicitly employ discarded spaces; this includes 
algorithms evoking the notion of tangent spaces \cite{Haegeman2011,Haegeman2013a,Haegeman2016,ZaunerStauber2018,Vanderstraeten2019} and generalizations thereof, as will be described later. 

To formulate the goals of this paper, we here briefly indicate how the nested subspaces mentioned above come about. Concrete
constructions follow in later sections.

An MPS $\ket{\Psi}$ written in canonical form is defined by a set of isometric tensors \cite{Schollwoeck2011}. The image space of an isometric tensor, its \textit{kept} space, is needed for the description of $\ket{\Psi}$. The orthogonal complement of the kept space, its \textit{discarded} space, is not needed for $\ket{\Psi}$ itself,
but for the description of changes of $\ket{\Psi}$ due to an update step, e.g.\ during variational optimization, time evolution, or the computation of excitations above the ground state. Any such change can be assigned to one of
the subspaces $\doubleVnsite$ in the nested hierarchy
\begin{align}
\label{eq:NestingVnsite}
\doubleVzerosite \subset \doubleVonesite \subset 
\doubleVtwosite \subset \dots \subset \doubleV^{\seLL \mathrm{s}} = 
\doubleV \, , 
 \end{align}
where $\doubleV$ is the  full Hilbert space of a system of $\eLL$ sites, 
  $\doubleVnsite$ the subspace spanned by all $n$-site\ ($n$s) variations of $\ket{\Psi}$, and  $\doubleVzerosite = \mathrm{span}\{\ket{\Psi}\}$ the one-dimensional space spanned by the reference MPS itself. The orthogonality of kept and
discarded spaces induces a partition of each $\doubleVnsite$ 
into nested orthogonal subspaces \cite{Haegeman2013,Hubig2018}, such that 
\begin{align}
\label{eq:DecomposeVnsite}
	\doubleVnsite = \oplus_{n'=0}^n \doubleV^{n' \perp} \, ,
\end{align}
where $\doubleVn$ is the subspace of $\doubleVnsite$ spanned by all irreducible $n$s variations not 
expressible through $n'$s variations with $n'< n$,
and $\doubleVzero=\doubleVzerosite$. In particular, the full Hilbert space
can be represented as 
$\doubleV = \oplus_{n=0}^{\seLL} \doubleV^{n \perp}$.
 
The subspaces defined above underlie, implicitly or explicitly, all MPS algorithms. $\doubleVonesite$ is the so-called tangent space 
 of $\ket{\Psi}$, i.e.\ the space of all one-site (\onesite) variations of $\ket{\Psi}$. It plays an explicit role in numerous recent MPS algorithms, such as TDVP time-evolution, or  the description of translationally invariant MPS and their excitations 
 \cite{Haegeman2013,ZaunerStauber2018,Vanderstraeten2019}. It also features implicitly in MPS algorithms formulated using \onesite\ update schemes, such as the \onesite\ formulation of DMRG \cite{Schollwoeck2011}, because \onesite\ updates
 explore states from $\doubleVonesite$.  Likewise, the space 
 $\doubleVtwosite$ implicitly underlies all \twosite\ MPS algorithms such as 
 \twosite\ DMRG ground state search, \twosite\ time-dependent DMRG (tDMRG),
 or \twosite\ TDVP, in that  \twosite\ updates explore states from $\doubleVtwosite$.
 Moreover, $\doubleVone$ and $\doubleVtwo$  
 are invoked explicitly when computing the \twosite\ energy variance,  
an error measure for MPS ground state searches introduced in Ref.~\cite{Hubig2018}.
Finally, $\doubleVnsite$ is implicitly invoked in MPS algorithms defining excited states of translationally invariant MPS through linear combinations of local excitations defined on
$n$ sites \cite{Haegeman2013a}.

The construction of a basis for $\doubleVnsite$ and $\doubleVn$ is well known
for $n\!=\!1$ \cite{Haegeman2016}, and for $\!n=\! 2$ 
it is outlined in Ref.~\cite{Hubig2018}. However, we are not aware of a general,
explicit construction for $n\!>\!2$, as needed, e.g., to compute the \nsite\ energy 
variance. Here, we explicitly construct projectors, $\Pnsite$ and $\Pn$, having $\doubleVnsite$ and $\doubleVn$ as images,
respectively. For $n=1$, this amounts to a construction of a basis
for the tangent space $\doubleVonesite$. More generally, our $\K,\! \D$ projector formalism used to construct $\Pnsite$ and $\Pn$
greatly facilitates the formulation of MPS algorithms that explicitly or implicitly employ discarded spaces. As an illustration, we derive an explicit expression for the $n$-site energy variance, generalizing the error measure  proposed in Ref.~\cite{Hubig2018}, and evaluate it numerically for a model with long-range hopping, the Haldane-Shastry model. We also show how the 
multiparticle \nsite\ excitations proposed in Ref.~\cite{Haegeman2013a}
are formulated in our scheme, and propose  a strategy
for computing them explicitly, for any $n$.

We expect that the $\K$,$\D$ projector formalism developed here will be particularly useful for improving the efficiency of  MPS algorithms
by incorporating information from $\doubleVn$ into suitably expanded
versions of $\doubleV^{(n'<n)\mathrm{s}}$ without fully computing
$\doubleVn$. For example, we have recently developed
a scheme, called controlled bond expansion, which incorporates \twosite\ information into \onesite\ updates for DMRG ground state search \cite{Gleis2022} and TDVP time evolution \cite{Li2022}, in a manner requiring only \onesite\ costs.

This paper is structured as follows. 
In Sec.~\ref{sec:Basics} we collect some well-known facts about MPSs, 
and formally define the associated kept and discarded spaces and corresponding projectors.  Section~\ref{sec:ConstructionOfPnPnPerpHierarchy}, the heart of this paper, describes the construction of the $\Pnsite$ and $\Pn$ projectors for general $n$. As applications of our projector formalism, we compute the \nsite\ energy variance of the Haldane-Shastry model in Sec.~\ref{eq:EnergyvarianceTangentProjector}, and describe the construction and computation of \nsite\ excitations in Sec.~\ref{sec:n-siteExcitations}.
We end with a brief outlook in Sec.~\ref{sec:Outlook}.

\section{MPS basics}%
\label{sec:Basics}

This section offers a concise, tutorial-style summary of MPS notation and the associated  diagrammatics.
Moreover, we formalize the notion of kept spaces, needed to describe
an MPS $\ket{\Psi}$, and discarded spaces, needed to describe changes to it
at specified sites. We also recapitulate the definition of local bond, \onesite\ and \twosite\ projectors routinely used in \onesite\ and \twosite\ MPS algorithms. 

\subsection{Basic MPS notation}
\label{sec:BasicMPSNotation}

Consider a quantum chain with sites labeled 
  $ \ell = 1, \ndots,  \eLL$.
Let each site be represented by a $d$-dimensional Hilbert space, $\doublev[\ell]$, with local basis states $|\sigma_\ell\rangle$, $\sigma_\ell = 1, \ndots, d$.
The full Hilbert space is $\doubleV = \! \prod_{\otimes \ell} 
\doublev[\ell] = \mathrm{span}\{ \ket{\vec{\sigma}}\}$,
with basis states $\ket{\vec{\sigma}} = 
\ket{\sigma_1} \ket{\sigma_2} \ncdots \ket{\sigma_\seLL}$. 
Any state $\ket{\Psi} = \ket{\vec{\sigma}} \Psi^{\vec{\sigma}}
\in \doubleV$
can be written as an open boundary MPS, with wavefunction of the form 
\begin{flalign}
\label{subeq:define-MPS}	
\Psi^{\vec{\sigma}} & =  
[M_1]^{\sigma_1}_{1 \alpha_1} [M_2]^{\sigma_2}_{\alpha_1 \alpha_2} \! \ncdots  
[M_\seLL]^{\sigma_\sseLL}_{\alpha_{\seLL-1} 1}  
\hspace{-1cm} & \\
& =  
     \raisebox{-5mm}{\includegraphics[width=0.733\linewidth]{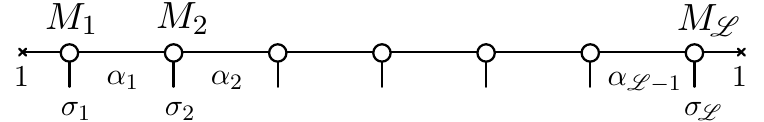}} 
     . \vspace{-2cm} & 
  \nonumber 
\end{flalign}
\noindent 
(This diagram depicts both the wavefunction $\Psi$ and the corresponding state $\ket{\Psi}$.) For clarity, we do not use ellipses in our MPS diagrams, but instead draw them for some small choice of $\eLL$, e.g.\ $\eLL = 7$ above.
Sums over repeated indices are implied throughout, and depicted
diagramatically by bonds. Each $M_\ell$ is a three-leg tensor with elements $[M_\ell]^{\sigma_\ell}_{\alpha_{\ell-1} \alpha_\ell}$. 
Its physical and virtual bond indices, $\sigma_\ell$ and $\alpha_{\ell-1}$, $\alpha_\ell$, have dimensions $d$ and $D_{\ell-1}$, $D_\ell$, respectively. 
The outermost bonds, to dummy sites represented by crosses, have $D_0\!=\! D_{\seLL} \!=\! 1$.
The bond dimensions $D_\ell$ are adjustable parameters, controlling the amount of entanglement an MPS can encode.
(In the literature, it is common practice to drop the subscript on $D_\ell$ for brevity, understanding that $D$ can nevertheless vary from bond to bond.) Likewise, a Hamiltonian acting within $\doubleV$, 
$\Hc = \ket{\vec{\sigma}} H^{\vec{\sigma}\vec{\sigma'}}
\bra{\vec{\sigma'}}$,  
can be expressed as an MPO, with 
\begin{flalign}
\label{subeq:define-H-MPO}	
H^{\vec{\sigma}\vec{\sigma'}}
& = [W_1]^{\sigma_1 \sigma'_1}_{1 \nu_1} [W_2]^{\sigma_2 \sigma'_2}_{\nu_1 \nu_2} \! \ncdots  
[W_\seLL]^{\sigma_{\sseLL} \sigma'_{\sseLL}}_{\nu_{\seLL-1} 1}  , 
\hspace{-1cm} & \\
& =  
     \raisebox{-4.4mm}{\includegraphics[width=0.743\linewidth]{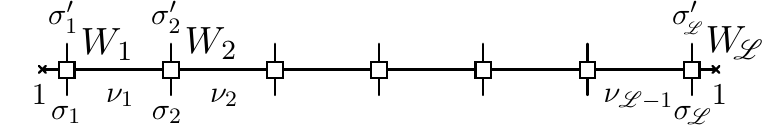}}  \, ,  \hspace{-1cm} &
    \nonumber
\end{flalign}
where the four-leg tensors $W_\ell$ have elements $[W_\ell]^{\sigma_\ell \sigma'_\ell}_{\nu_{\ell-1} \nu_\ell}$, and the virtual bond indices
$\nu_\ell$ have dimensions $w_\ell$.

Any MPS wavefunction can be brought into canonical form w.r.t.\ 
an ``orthogonality center'' at site $\ell \in [1, \eLL]$,
or w.r.t.\ bond $\ell$ connecting sites $\ell$ and $\ellplusone$,
\begin{align}
	\Psi^{\vec{\sigma}} & =  
	\raisebox{-5.6mm}{\includegraphics[width=0.733\linewidth]{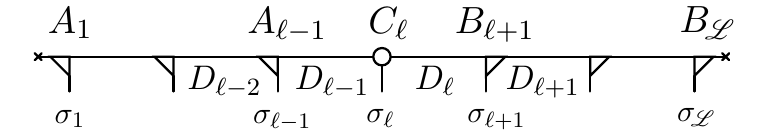}} ,
     \label{eq:canonical}
\end{align}
where we indicated some of the bond dimensions. Here, $A_\tell$ and $B_{\tellp}$ (with $1 \!\le \! \tell \! < \! \ell \! < \! \tellp \! \le \! \eLL$) 
satisfy the relations
\begin{flalign} 
\nonumber
& \big[A_\tell^\dagger\big]^{\sigma}_{\alpha\bar \alpha} 
\big[A^\pdag_\tell\big]^{\sigma}_{\bar \alpha \alpha'}
  \! = \! \big[\doubleIk[\tell]\big]_{\alpha\alpha'} , 
 \quad \!
 \big[B_\tellp^\pdag\big]^{\sigma}_{\alpha\bar \alpha} 
\big[B^\dagger_\tellp\big]^{\sigma}_{\bar \alpha \alpha'}
 \!  = \! \big[\doubleIk[\tellp-1]\big]_{\alpha\alpha'} ,  & 
\\
 & 
 \raisebox{-6.5mm}{
 \includegraphics[width=0.9\linewidth]{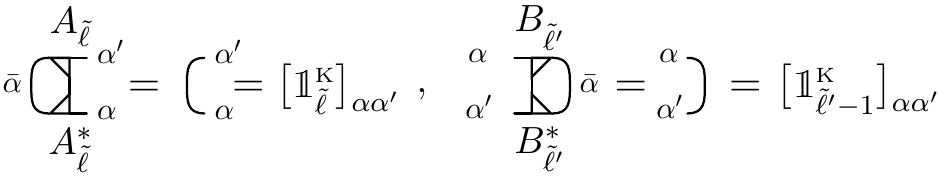}} \; , 
 \hspace{-1cm} &
    \label{eq:isometry}
\end{flalign}
or $A_\tell^\dagger A^\pdag_\tell = \doubleIk[\tell]$, 
$B_\tellp^\pdag B^\dagger_\tellp = \doubleIk[\tellp-1]$ 
for short, where  
$\doubleIk[\tell]$ denotes a $D_\tell\times\! D_\tell$ unit matrix.
(The superscript $\K$ stands for ``kept'', for reasons explained below.)
The open triangles representing $A_\tell$ and $B_\tellp$ are oriented
such that their diagonals face left or right, respectively.
The orthogonality center can be shifted left or right by using
singular value decomposition (SVD) to express it as 
$C_\ell = U_{\ell-1} S_{\ell-1} B_\ell$ or 
$C_\ell = A_\ell S_{\ell} V^\dag_\ell$:
\begin{flalign}
&	\label{eq:shift-orthogonality-center}
 \raisebox{-8mm}{
\includegraphics[width=0.966\linewidth]{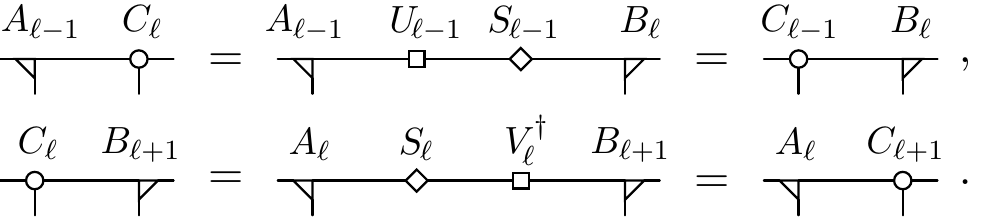}}
\hspace{-1cm} 
& & 
\end{flalign}
Here $U_{\ell-1}$, $V^\dag_\ell$, $S_{\ell-1}$, $S_\ell$ are square matrices, the former two unitary, the latter two diagonal and containing SVD singular values.
(Shifting can be combined with truncation, if desired, by discarding some small singular values and correspondingly reducing the bond dimension.)
By renaming $V^\dag_\ell B_\ellplusone$ as $B_\ellplusone$
and defining $\Lambda_\ell = S_\ell$, 
we can also express 
$\Psi^{\vec{\sigma}}$ in ``bond-canonical" form w.r.t.\ 
bond $\ell$:
\begin{align}
	\label{eq:bond-canonical-form}
& \raisebox{-3mm}{
\includegraphics[width=0.78\linewidth]{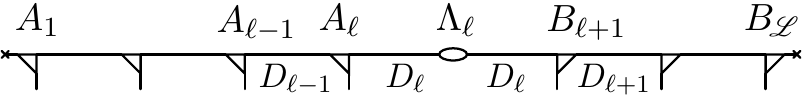}} \, .
\end{align}
The fact that the same MPS can be written in many different but equivalent ways reflects the gauge 
freedom of MPS representations.

\subsection{Kept spaces}
\label{sec:KeptSpaces}

Given an MPS $\ket{\Psi}$ in canonical form,  its constituent tensors
can be used to define a set of state spaces defined on parts of the chain, and a sequence of isometric maps between these state spaces. Let us make this explicit to reveal the underlying structures. 

The $A_\tell$ tensors for sites $1$ to $\tell$
can be used to define a set of left  \textit{kept} ($\K$) states 
$\ket{\Psik_{\tell \alpha}}$, and the $B_\tellp$ tensors for sites $\tellp$ to $\eLL$ 
can be used to define right $\K$ states $\ket{\Phik_{\tellp \alpha'}}$, 
with wavefunctions of the form 
\begin{align}
 \Psik_{\tell \alpha} =   \raisebox{-2.2mm}{
 \includegraphics[width=0.233\linewidth]{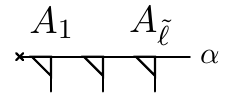}}, 
 \quad
 \Phik_{\tellp \alpha'} =   \raisebox{-2.2mm}{
 \includegraphics[width=0.233\linewidth]{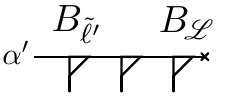}} 
     \label{eq:OrthonormalStatesKept}.
\end{align}
These states are called \textit{kept}, since they are building blocks of $\ket{\Psi}$.
Their spans define left and right $\K$ spaces, 
\begin{align}
	\label{eq:defineStateSpaces}
\doubleVk[\tell] & = \mathrm{span}\{ \ket{\Psik_{\tell \alpha}} \}
 \; \, \subset  \doublev[1] \otimes \ndots \otimes 
\doublev[\tell] \, , \\ 
\doubleWk[\tellp] & = 
\mathrm{span}\{ \ket{\Phik_{\tellp \alpha'}} \}  \subset 
\doublev[\tellp] \otimes \ndots \otimes \doublev[\sseLL] \, ,
\end{align}
of dimension $D_\tell$ and $D_{\tellp-1}$, respectively.
The dummy sites $0$ and $\eLL+1$  are represented by
one-dimensional spaces, $\doubleVk[0]$ and $\doubleWk[\seLL+1]$. 

Each $A_\tell$ and $B_\tellp$ tensor defines an isometric map, from 
a \textit{parent} ($\P$) space involving a direct product of a $\K$ space and a local space, to an adjacent $\K$ space:
\begin{align}
\nonumber
	A_\tell \! : \! \doubleVk[\tell-1] \! \otimes \! \doublev[\tell] & \to 
	\doubleVk[\tell], 
&
	\ket{\Psik_{\tell-1, \alpha}} \ket{\sigma_\tell} 
	\bigl[A_\tell]^{\sigma_\tell}_{\alpha \alpha'}
	& = \ket{\Psik_{\tell \alpha'}} , 
\\
\nonumber
	B_\tellp \! : \! \doublev[\tellp] \! \otimes \! \doubleWk[\tellp+1] \!  & \to 
	\doubleWk[\tellp], 
	& \!
	\bigl[B_\tellp]^{\sigma_\tellp}_{\alpha \alpha'} 
	\ket{\sigma_\tellp}  \ket{\Phik_{\tellp+1, \alpha'}} 
	& = \ket{\Phik_{\tellp \alpha}}  .
\end{align}
(To connect sites 1 and $\eLL$ to their neighboring dummy
sites,  we define $\Psik_{0, 1} \!=\! 1$, $\Phik_{\seLL+1, 1} \!=\! 1$.)
We orient the triangles depicting $A_\tell$ and $B_\tellp$
such that equal-length legs point to parent spaces and 90-degree angles  to kept spaces. The dimensions of left or right
kept and parent spaces satisfy $D_\tell \le D_{\tell-1}d $ or 
$D_{\tellp-1} \le d D_{\tellp} $, respectively. If a kept space is smaller than its parent space, it has an orthogonal complement, called \textit{discarded} ($\D$) space, discussed 
in Sec.~\ref{sec:OrthogonalProjectors} below.  
The fact that the maps $A_\tell$ and $B_\tellp$ are \textit{isometries}
follows from  Eqs.~\eqref{eq:isometry}. These ensure that the 
left and right $\K$
basis states form orthonormal sets, 
\begin{flalign}
\nonumber 
& \quad \;  \ovl{\Psik_{\tell \alpha}}{\Psik_{\tell \alpha'}} 
	 =  
	\bigl[\doubleI[\tell]^\ks\bigr]_{\alpha  \alpha'} \, ,  \,  \quad  
	\ovl{\Phik_{\tellp \alpha}}{\Phik_{\tellp \alpha'}}  =  
	\bigl[\doubleI[\tellp-1]^\ks \bigr]_{\alpha \alpha'} \, , 
	 &
\\[2mm]	
 & \quad \, \hspace{-2mm} \raisebox{-6.25mm}{
 \includegraphics[width=0.8\linewidth]{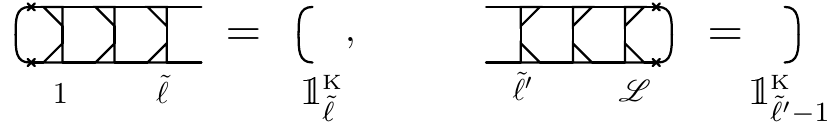}} . 
 \hspace{-1cm} &
    \label{eq:orthonormalbases}
\end{flalign}
The basis states can be used to build projectors 
onto the left or right $\K$  spaces $\doubleVk[\tell]$
or $\doubleWk[\tellp]$, depicted as
\begin{subequations}
\label{subeq:PQprojectors}
  \begin{align}
 \Pk_\tell = 
 \sum_\alpha \prj{\Psik_{\tell \alpha}}  & \; = 
 \raisebox{-4.2mm}{
 \includegraphics[width=0.167\linewidth]{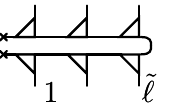}} \; ,
\\[-0.5mm] 
\Qk_\tellp  = 
\sum_{\alpha'}	\prj{\Phik_{\tellp \alpha'}} 
 & \; = 
\raisebox{-4.2mm}{
 \includegraphics[width=0.167\linewidth]{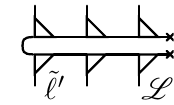}} \; ,
\end{align}
\end{subequations}
with $\Pk_0 \!=\! 1$, $\Qk_{\seLL+1} \!=\! 1$, and
$(\Pk_\tell)^2 \!=\! \Pk_\tell$, $(\Qk_\tellp)^2 \!=\! \Qk_\tellp$:
\begin{align} \raisebox{-0.5\height}{
 \includegraphics[width=0.8\linewidth]{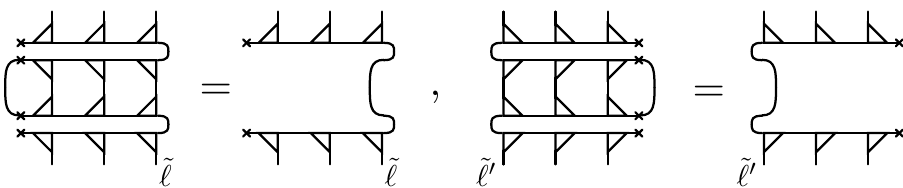}} \; . 
    \label{eq:PSquaredQSquared}
\end{align}

\subsection{Bond, 1s and 2s projectors}
\label{sec:ProjectedHamiltonians}

The above projectors can, in turn, be used to construct bond, \onesite\ and \twosite\ projectors acting on the full chain, 
\begin{subequations}
	\label{eq:0-1-two-site-projectors}
\begin{flalign}
\label{eq:P0}
\Pbond_{\ell}  = \Pk_{\ell} \!\otimes\! \Qk_{\ellplusone} 
 & = \!\!
\raisebox{-5.3mm}{
 \includegraphics[width=0.317\linewidth]{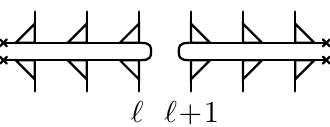}} \; ,
 \hspace{-1cm} & 
\\
 \label{eq:P1}
 \Ponesite_{\ell}  = \Pk_{\ell-1} \!\otimes\! \doubleI[d] \!\otimes\! \Qk_{\ellplusone} 
  & = \!\!
\raisebox{-5.3mm}{
 \includegraphics[width=0.317\linewidth]{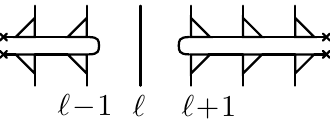}} \; ,
 \hspace{-1cm}  & 
\\
\label{eq:P2}
\Ptwosite_{\ell}  = \Pk_{\ellminusone} \!\otimes\!  \doubleI[d] \!\otimes\!  \doubleI[d] \!\otimes\!  \Qk_{\ellplustwo} 
& = \!\!
\raisebox{-5.3mm}{
 \includegraphics[width=0.317\linewidth]{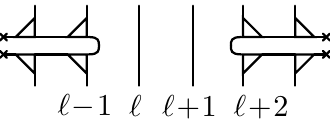}}\; ,
\hspace{-1cm} & 
\end{flalign}
\end{subequations}
defined for 
$\ell \!\in\! [0,\eLL]$,
$\ell \!\in\! [1,\eLL]$ and 
$\ell \!\in\!  [1,\eLL\!-\!1]$, respec\-tively.
They mutually commute and satisfy $(\Pc^\xs_\ell)^2 \!=\! \Pc^\xs_\ell$,
as follows from \Eqs{eq:orthonormalbases} and \eqref{eq:PSquaredQSquared}.
For example:  
\begin{flalign} 
\nonumber
& (\Pbond_\ell)^2=
\raisebox{-0.475\height}{\includegraphics[width=0.7\linewidth]{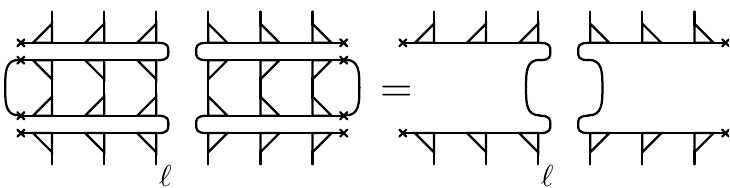}} 
\!\! = \Pbond_\ell   .
\hspace{-1cm} &
\end{flalign}
The projectors $\Pbond$, $\Ponesite$ and $\Ptwosite$
map the full $\doubleV$ into the subspaces 
$\doubleVk[\ell] \otimes \doubleWk[\ellplusone]$,
$\doubleVk[\ell-1] \otimes \doublev[\ell] \otimes \doubleWk[\ellplusone]$
and $\doubleVk[\ell-1]  \otimes \doublev[\ell] \otimes 
\doublev[\ellplusone] \otimes \doubleWk[\ell+2]$. These spaces offer three 
equivalent representations of the same state $\ket{\Psi}$, in bond-, \onesite- or \twosite-canonical form, 
\begin{subequations}
	\label{subeq:canonicalforms}
\begin{flalign}
	\label{eq:bond-representation}
	\ket{\Psi} & = 
	\ket{\Psik_{\ell \alpha}} \ket{\Phik_{\ell+\hspace{-0.2mm}1, \alpha'}} 
    \bigl[\psi^\mbond_\ell \bigr]_{\alpha \alpha'}
\\ 
\label{eq:one-site-representation}
&  = 
	\ket{\Psik_{\ell\hspace{-0.2mm}-\hspace{-0.2mm}1,  \alpha}} 
	\ket{\sigma_\ell} \ket{\Phik_{\ell+\hspace{-0.2mm}1, \alpha'}} 
	\bigl[\psi^\monesite_\ell \bigr]^{\sigma_{\ell}}_{\alpha \alpha'}
\\ 
\label{eq:two-site-representation}
& = 
	\ket{\Psik_{\ellminusone, \alpha}} 
	\ket{\sigma_\ell} 
	\ket{\sigma_{\ellplusone}} 
	\ket{\Phik_{\ellplustwo, \alpha'}\!}  
		\bigl[\psi^\mtwosite_\ell \bigr]^{\sigma_{\ell}\sigma_{\ellplusone}}_{\alpha \alpha'} \, , 
 \\	
 	\psi^\mbond_\ell & = \Lambda_\ell , \quad 
 	\psi^\monesite_\ell = C_\ell , \quad 
 	\psi^\mtwosite_\ell = A_{\ell} \Lambda_\ell B_{\ellplusone} . 
\end{flalign}
\end{subequations}
These forms emphasize the tensors describing bond $\ell$, site $\ell$ or sites $(\ell, \ell\!+\!1)$ and the bond in between, respectively.
For example, \Eqs{eq:bond-representation} and \eqref{eq:one-site-representation} are depicted as
\begin{align}
\nonumber
	\Psi & =  
	\raisebox{-7.6mm}{\includegraphics[width=0.906\linewidth]{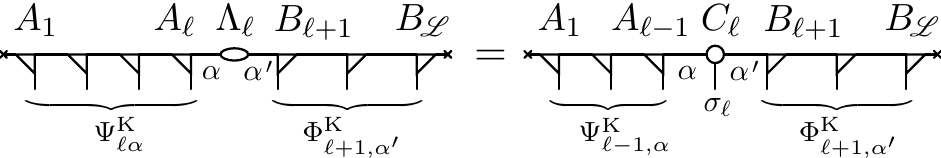}} \, .
\end{align}

The projections of the Hamiltonian into these spaces, 
$\Hc^\mxsite_\ell = \Pc^\mxsite_\ell \Hc \Pc^\mxsite_\ell$, 
have matrix elements of the form
\begin{flalign}
\label{eq:H1}
& H^\mbond_\ell \! = \!\!
\raisebox{-5.3mm}{
 \includegraphics[width=0.116\linewidth]{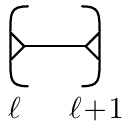}} \! ,
\;\;\; 
	H^\monesite_\ell \! = \!\!\!\!
\raisebox{-5.3mm}{
 \includegraphics[width=0.15\linewidth]{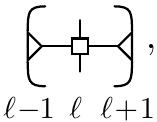}} 
\;\;\; 
H^\mtwosite_\ell \! = \!\!\!\!
\raisebox{-5.3mm}{
 \includegraphics[width=0.216\linewidth]{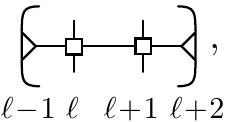}} 
\hspace{-5mm} &
\end{flalign}
with left and right environments for sites $\ell\pm 1$ given by
\begin{subequations}
	\label{eq:L-R-environments}
\begin{align}
	\label{eq:L-environments}
L_\ell & =  \raisebox{-5.3mm}{
 \includegraphics[width=0.45\linewidth]{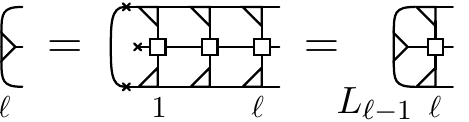}} \, ,
\\ 
	\label{eq:R-environments}
R_\ell & = \raisebox{-5.3mm}{
 \includegraphics[width=0.5\linewidth]{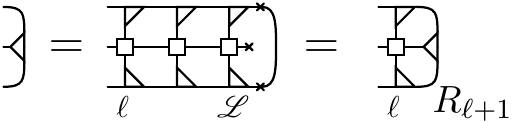}} \, .
\end{align}
\end{subequations}
Here the first equalities define $L_\ell$ and $R_\ell$,
the second equalities show how they can be computed recursively,
starting from $L_0=1$, $R_{\seLL+1}= 1$.
The open triangles on $L_\ell$ and $R_\ell$ signify that they
are computed using left- or right-normalized $A$ or $B$ tensors.

The above matrix elements are standard ingredients in numerous MPS algorithms. To give a specific example, we briefly recall their role in DMRG ground state searches. These
seek approximate ground state solutions to $\Hc \ket{\Psi} = E \ket{\Psi}$ through a sequence of local optimization steps. Focusing on bond $\ell$, or site $\ell$, or sites $(\ell, \ell\!+\!1)$, 
one 
updates $\Lambda_\ell$, or $C_\ell$, or 
$A_{\ell} \Lambda_\ell B_{\ellplusone}$, 
by finding the ground state solution of, respectively, 
\begin{subequations}
\label{subeq:one-site-two-site-SchroedingerEquation}	
\begin{flalign}
\label{eq:bond-SchroedingerEquation}	
		(H_\ell^\mbond \!-\! E) \psi^\mbond_\ell 
		& =  0 \,  , \qquad \;  
\raisebox{-6.3mm}{
 \includegraphics[width=0.3\linewidth]{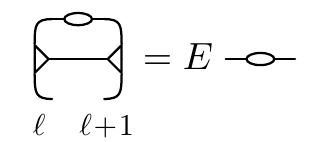}} ,  \hspace{-1cm} & 
\\  
	\label{eq:one-site-Schroedinger}
		(H_\ell^\monesite \!-\! E) \psi^\monesite_\ell 
		& =  0  \, , \qquad \; 
\raisebox{-5.3mm}{
 \includegraphics[width=0.3\linewidth]{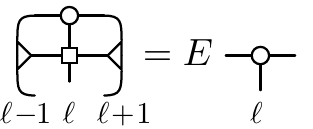}} , \hspace{-1cm} & 
\\  
	\label{eq:two-site-Schroedinger}
\quad      (H_\ell^\mtwosite \!-\! E) \psi^\mtwosite_\ell
     & = 0 \, , \;\; 
	\raisebox{-5.3mm}{
 \includegraphics[width=0.44\linewidth]{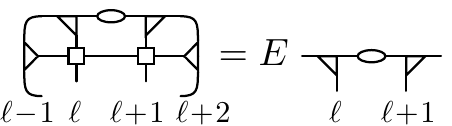}} . \hspace{-1cm} &
\end{flalign}
\end{subequations}
One then uses \Eq{eq:shift-orthogonality-center} to shift the orthogonality center to the neighboring bond or site, optimizes it, and sweeps back and forth through the chain until the ground state energy has converged.
These three schemes are known as \zerosite\ or bond DMRG, \onesite\ and \twosite\ DMRG, respectively. They differ regarding their flexibility for increasing (``expanding'') virtual bond dimensions, which increases the size of the variational space and hence the accuracy of the converged ground state energy. \zerosite\ and \onesite\ DMRG offer no way of doing this, because the tensors $\Lambda_\ell$ or $C_\ell$ have the same dimensions after the update as before. 
By contrast, \twosite\ DMRG does offer a way of expanding
bond dimensions: the bonds connecting 
the updated tensors $A_{\ell}$, $\Lambda_\ell$ and $B_{\ellplusone}$
have dimensions $d \, \mathrm{min}(D_{\ell- 1},D_{\ellplusone})$, which is $\ge D_\ell$; one may thus expand bond $\ell$ by retaining more than
$D_\ell$ singular values in $\Lambda_\ell$. 
However, this comes at a price. The numerical cost is  $\mathcal{O}(D^3 d^2 w)$ for 
applying $H^\mtwosite$ to $\psi^\mtwosite$ during the iterative solution of the eigenvalue problem Eq.~\eqref{eq:two-site-Schroedinger},
and $\mathcal{O}(D^3 d^3)$ for SVDing the resulting eigenstate
to identify the updated $A$, $\Lambda$, and $B$. By contrast, for \onesite\ DMRG the costs are lower: $\mathcal{O}(D^3 d w)$ for applying $H^\monesite$ to $C$, and
$\mathcal{O}(D^3 d)$ for SVDing $C$ to shift to the next site.
Various schemes have
been proposed for achieving \twosite\ accuracy at \onesite\ costs, 
see Refs.~\cite{White2005,Hubig2015,Gleis2022}.

\subsection{Discarded spaces}
\label{sec:OrthogonalProjectors}

In this section, we define discarded spaces as the orthogonal complements of kept spaces, and introduce their corresponding isometries and discarded space projectors.
As mentioned above, the kept spaces $\doubleVk[\tell]$ and 
$\doubleWk[\tellp]$ have dimensions smaller than
the parent spaces $\doubleVk[\tell-1] \! \otimes \! \doublev[\tell]$
and $\doublev[\tellp] \! \otimes \!  \doubleWk[\tellp+1]$ from which they are constructed. Their orthogonal complements are the above-mentioned
discarded spaces, to be denoted $\doubleVd[\tell]$ and  $\doubleWd[\tellp]$,
respectively, of dimension 
$\Db{}_{\tell}^{A} = D_{\tell-1} d- D_\tell$ and 
$\Db{}_{\tellp}^{B} = D_{\tellp}  d-D_{\tellp-1}$.
By definition, $\mathrm{span}\{\doubleVk[\tell], \doubleVd[\tell]\} $
and $\mathrm{span}\{\doubleWk[\tellp], \doubleWd[\tellp] \}$ yield
the full parent spaces, respectively. Let
$\Ab_\tell$ and $\Bb_\tellp$ be isometries from 
the parent to the discarded spaces, 
\begin{align}
\nonumber
\Ab_\tell \! : \! \doubleVk[\tell-1] \! \otimes \! \doublev[\tell]
& \to  \doubleVd[\tell], 
&
	\ket{\Psik_{\tell-1, \alpha}} \ket{\sigma_\tell} 
	\bigl[\Ab_\tell]^{\sigma_\tell}_{\alpha \alpha'}
	& = \ket{\Psid_{\tell \alpha'}} , 
\\
\nonumber
\Bb_\tellp \! : \! \doublev[\tellp] \! \otimes \!  \doubleWk[\tellp+1]  
& \to \doubleWd[\tellp] , 
	& \!
	\bigl[\Bb_\tellp]^{\sigma_\tellp}_{\alpha \alpha'} 
	\ket{\sigma_\tellp}  \ket{\Phik_{\tellp+1, \alpha'}} 
	& = \ket{\Phid_{\tellp \alpha}}  .
\end{align} 
Then $A_\tell \oplus \Ab_\tell $ and $B_\tellp \oplus \Bb_\tellp$ are unitary maps on the parent spaces, and 
Eq.~\eqref{eq:isometry} is complemented by relations
expressing orthonormality and completeness: 
\begin{flalign}
\nonumber
&	\Ab_\tell^\dag \Ab^\pdag_\tell   = \doubleId[\tell] \, , \;\; 
    A_\tell^\dag \Ab{}^\pdag_\tell   = 0 \, ,  \;\; 
	\Bb_\tellp^\pdag \Bb^\dag_\tellp  = \doubleId[\tellp-1] \, , \;\;
	\Bb{}^\pdag_\tellp B^\dag_\tellp   = 0 \, ,  &
\\[0.5mm]	\label{eq:orthogonality}
& \raisebox{-4mm}{
\includegraphics[width=0.89\linewidth]{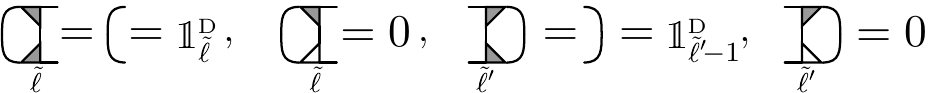}} , 
  \hspace{-1cm} &
\\[1mm] \label{eq:completeness-1}
& A^\pdag_\tell A^\dag_\tell \!+\! \Ab^\pdag_\tell \Ab^\dag_\tell = 
\doubleIp[\tell]   , 
\quad  \; 
B^\dag_\tellp B^\pdag_\tellp \!+\! \Bb^\dag_\tellp \Bb^\pdag_\tellp  = 
\doubleIp[\tellp-1]   ,  \hspace{-2cm} &
\nonumber \\[0mm]	 
 &  \raisebox{-3.5mm}{
 \includegraphics[width=0.886\linewidth]{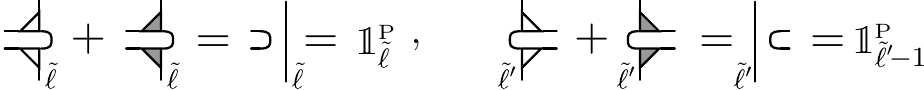}} \, .
 \hspace{-2cm} &
\end{flalign}
Here, left- or right-oriented grey triangles denote the complements $\Ab_\tell$ and $\Bb_\tellp$ associated with discarded spaces. The orthogonality relations
\eqref{eq:isometry} and 
\eqref{eq:orthogonality} state that $\K$ meeting $\K$ or $\D$ meeting $\D$ yield unity,
whereas $\K$ meeting $\D$ yields zero. We will use them often below.
For the completeness relations \eqref{eq:completeness-1},
$\doubleIp[\tell] \!=\! \doubleIk[\tell-1] \!\otimes\! \doubleI[d]$ and
$\doubleIp[\tellp-1] \!=\! \doubleI[d] \!\otimes\! \doubleIk[\tellp]$
are identity matrices on the parent spaces,
with $\doubleI[d]$ a $d \!\times\! d$ unit matrix. 
In numerical practice, it desirable to avoid the
explicit computation of $\Ab^\pdag_\tell \Ab^\dag_\tell$ or $\Bb^\dag_\tellp \Bb^\pdag_\tellp$, since these are huge objects. Instead, one can always
use \Eq{eq:completeness-1} to express them as
$\doubleIp[\tell] \!- \!A^\pdag_\tell A^\dag_\tell$ or $\doubleIp[\tellp\!-\!1] \!\!- \!B^\dag_\tellp B^\pdag_\tellp$. 

Equations~\eqref{eq:completeness-1} imply additional identities that will likewise be useful below:
\begin{subequations}
\label{eq:OneAndTwoSiteCompleteness}
\begin{flalign}
\label{eq:OneSiteCompletness}
&  \! \raisebox{-4.5mm}{
 \includegraphics[width=0.733\linewidth]{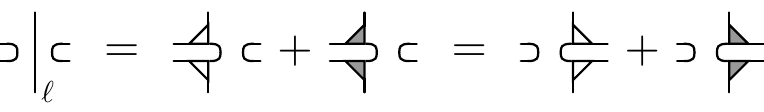}}\, , \hspace{-1cm} & 
\\
\label{eq:TwoSiteCompleteness}
& \! \raisebox{-4.5mm}{
 \includegraphics[width=0.85\linewidth]{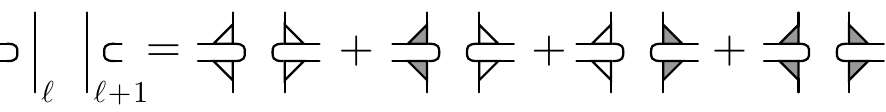}}
 \; , \hspace{-1cm} &
\\ 
\label{eq:DDTwoSiteCompleteness}
& \! \raisebox{-4.5mm}{
 \includegraphics[width=0.85\linewidth]{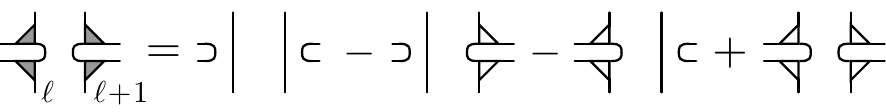}}
 \; .  \hspace{-1cm} &
\end{flalign}
\end{subequations}
The first two lines can be used to express \onesite\ or \twosite\ projectors
through bond projectors, as elaborated below. The third line follows from 
the first two.
The two equivalent forms on the right of \Eq{eq:OneSiteCompletness} 
arise from combining the physical state space of site $\ell$ with virtual state spaces on either the left or the right, yielding either left- or right-normalized parent spaces.

In complete analogy to \Eqs{eq:OrthonormalStatesKept} 
to \eqref{subeq:PQprojectors}, the complement isometries can be used to define 
orthonormal bases states for the left and right discarded spaces
$\doubleVd[\tell]$ and $\doubleWd[\tellp]$,  
\begin{flalign}
 & \Psid_{\tell \alpha} =   \raisebox{-2.2mm}{
 \includegraphics[width=0.233\linewidth]{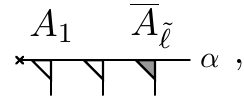}} 
 \quad
 \Phid_{\tellp \alpha'} =   \raisebox{-2.2mm}{
 \includegraphics[width=0.233\linewidth]{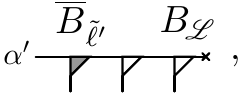}} 
     \label{eq:OrthonormalStates}
\hspace{-1cm} &
\end{flalign}
satisfying the orthonormality relations
\begin{subequations}
\label{eq:OrthonormalityDDDKKD}
\begin{flalign} 
& \quad \, \hspace{-2mm} \raisebox{-6.25mm}{
 \includegraphics[width=0.8\linewidth]{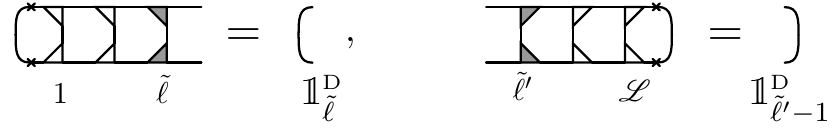}} , 
 \hspace{-1cm} &
\\
& \quad \, \hspace{-2mm} \raisebox{-6.85mm}{
 \includegraphics[width=0.8\linewidth]{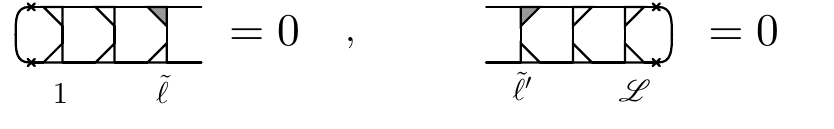}} . 
 \hspace{-1cm} &
\end{flalign}
\end{subequations}
The corresponding projectors are defined as 
\begin{align}
\label{eq:PQprojectorsDiscarded}
 \Pd_\tell = 
 \sum_\alpha \prj{\Psid_{\tell \alpha}}  & 
     \;= \raisebox{-4.2mm}{
 \includegraphics[width=0.167\linewidth]{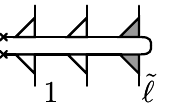}} \; ,
\\[-0.5mm] 
\Qd_\tellp  = 
\sum_{\alpha'}	\prj{\Phid_{\tellp \alpha'}} 
 & \; = 
\raisebox{-4.2mm}{
 \includegraphics[width=0.167\linewidth]{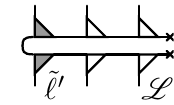}} \, ,
\end{align}
with $\Pd_0 \!=\!\Qd_{\seLL+1}\!=\!0$.
They obey orthonormality relations,  
\begin{align}
\label{eq:PxxbOrthonormality}
\Px_\tell \Pxb_\tell = 
\delta^{\xs \xsb} \Px_\tell \, , 
\qquad
\Qx_\tellp \Qxb_\tellp = 
\delta^{\xs \xsb} \Qxb_\tellp \, ,
\end{align}
where here and henceforth, $\X, \overline \X \in \{ \K, \D\}$.
Moreover, Eq.~\eqref{eq:completeness-1} implies the completeness relations
\begin{flalign}
\label{eq:PQ_recursion}
& \Pk_\tell + \Pd_\tell = \Pk_{\tell-1} \otimes \doubleI[d] \; , \;
\Qk_\tellp + \Qd_\tellp =  \doubleI[d] \otimes \Qk_{\tellp+1} \, ,
\hspace{-1cm} &
\end{flalign}
stating that the kept and discarded projectors of a given site 
together form a projector for their parent space. 
These will play a crucial role in subsequent sections.

To conclude this section, we apply the projector identity \eqref{eq:TwoSiteCompleteness}
to the open legs of the state $\Hc^\mtwosite_\ell \psi^\mtwosite_\ell$ appearing
in the \twosite\ Schr\"odinger \eqref{eq:two-site-Schroedinger}. We obtain:
\begin{flalign}
	\label{eq:H2-expressed-through-zero-site-terms}
&	\! \! \raisebox{-1mm}{ \includegraphics[width=0.916\linewidth]{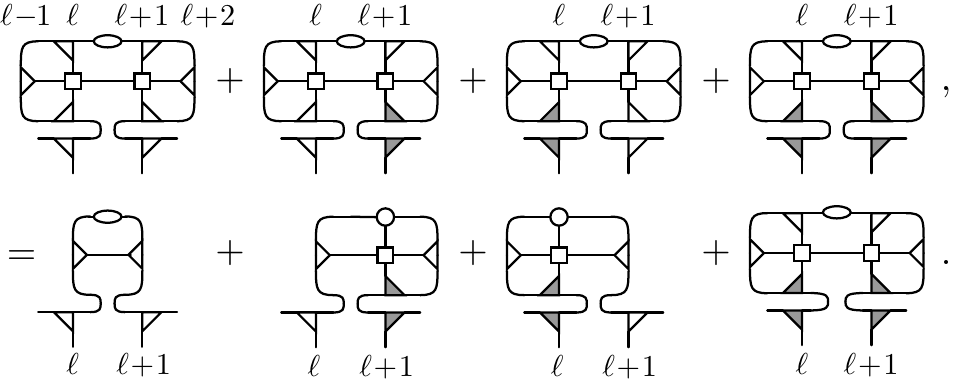}} \hspace{-1cm} &
\end{flalign}
If only the first term is retained, the \twosite\ Schr\"odinger Eq.~\eqref{eq:two-site-Schroedinger} reduces to the bond Schr\"odinger Eq.~\eqref{eq:bond-SchroedingerEquation}, sandwiched between $A_\ell$ and $B_{\ellplusone}$:
\begin{subequations}
\label{subeq:bond1s2sProjectedSchroedinger}	
\begin{align}
\label{eq:two-site-projected-to-bond-SchroedingerEquation}	
		A_\ell (H_\ell^\mbond \!-\! E) \Lambda_\ell B_{\ellplusone} & =  0  \, .  
\end{align}
The first term together with the second or third therm reduces to the \onesite\ Schr\"odinger Eq.~\eqref{eq:one-site-Schroedinger} for sites $\ellplusone$ or $\ell$, left- or right-contracted with  $A_\ell$ and $B_{\ellplusone}$, respectively:
\begin{align}
A_\ell (H_\ellplusone^\monesite \!-\! E) C_\ellplusone & =  0 
\\
(H_\ell^\monesite \!-\! E) C_\ell B_{\ellplusone} & =  0
\, .
\end{align}
All four terms together of course give the full \twosite\ Schr\"odinger Eq.~\eqref{eq:two-site-Schroedinger},
\begin{equation}
(H_\ell^\mtwosite \!-\! E)  A_\ell \Lambda_\ell B_{\ellplusone} = 0 \, .
\end{equation}
\end{subequations}
Evidently, the fourth term in \Eq{eq:H2-expressed-through-zero-site-terms}, 
involving a $\DD$ projector pair, is 
beyond the reach of \onesite\ schemes. 
A strategy for nevertheless computing its most important contributions with \onesite\ costs, called controlled bond expansion,
 has recently been formulated by us in Ref.~\cite{Gleis2022}.

\section{Construction of $\Pnsite$ and $\Pn$}
\label{sec:ConstructionOfPnPnPerpHierarchy}

As discussed in the introduction, each site of an MPS $\ket{\Psi}$ induces a splitting of the local
Hilbert space into $\K$ and $\D$ sectors. This induces a partition
of the full vector space $\doubleV$ into intricately nested orthogonal subspaces~\cite{Hubig2018}. It is useful to identify orthogonal projectors for these subspaces. Gauge invariance---the existence of many equivalent representations of $\ket{\Psi}$---makes this a nontrivial task. It can be accomplished systematically by Gram-Schmidt orthogonalization, formulated in projector language. The following three sections are devoted to this endeavor.

In the present section, we define a set of projectors, $\Pxxb_{\ell \ellb}$,
$\X,\Xb \in \{ \K,\D\}$, involving
kept and/or discarded sectors at sites $\ell,\ellb$. These serve as building blocks for all projectors 
introduced thereafter. Then, in Sec.~\ref{sec:GeneralizedBondProjectors}, 
we define generalized \textit{local} $n$-site (\nsite) projectors, 
$\Pnsite_\ell$,  
describing variations of $\ket{\Psi}$ involving up to $n$ contiguous sites. In Sec.~\ref{sec:ProjectorsnSiteVariations}, we add them up to obtain \textit{global} \nsite\ projectors, $\Pnsite$; and in Sec.~\ref{sec:OrthogonalGlobalnsProjectors} we  orthogonalize these to obtain 
\textit{irreducible} global \nsite\ projectors, $\Pn$, not expressible through combinations of variations on subsets of $n'< n$ sites. 
They are useful for various purposes, 
including the computation of the energy variance \cite{Hubig2018}, and the formulation
of MPS algorithms based on the notion of tangent spaces \cite{Lubich2015a,Haegeman2016,ZaunerStauber2018,Vanderstraeten2019,Li2022} and generalizations thereof.
Throughout, we concisely summarize the properties of the various projectors encountered along the way.

\subsection{Projectors for kept and discarded sectors, $\Pxxb_{\ell \ellb}$}
\label{sec:PxxbProjectors}

We start by introducing kept and discarded space projectors defined on the full Hilbert space $\doubleV$. To this end, we supplement $\Pc^\xs_\ell$ and $\Qc^\xs_\ell$ by
right or left environments ($\E$) comprising the entire rest of the chain, and define
\begin{align}
\Pxe_{\ell} &= \Px_{\ell}\otimes\doubleI[d]^{\otimes \, \scripteLL-\ell}
\, ,\quad &
\Pex_{\ell} &= \doubleI[d]^{\otimes \ell-1}\otimes \Qx_{\ell} \, , 
\\
\nonumber \Pke_{\ell} & = 
\raisebox{-0.45\height}{\includegraphics[width=0.317\linewidth]{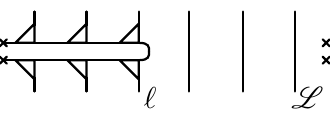}}  \, , 
& 
\Pek_{\ell} & = 
\raisebox{-0.45\height}{\includegraphics[width=0.317\linewidth]{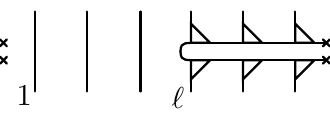}} 
 \, , 
\\[-1.5mm]
\nonumber \Pde_{\ell} & = 
\raisebox{-0.45\height}{\includegraphics[width=0.317\linewidth]{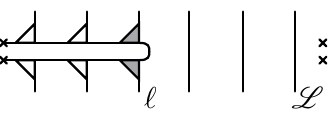}} 
 \,  , 
& 
\Ped_{\ell} & = 
\raisebox{-0.45\height}{\includegraphics[width=0.317\linewidth]{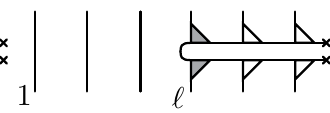}} 
 \, . \quad
\end{align}
with $\ell \in [0,\eLL]$ for $\Pxe_\ell$ and $\ell \in [1,\eLL+1]$
for $\Pex_\ell$.
Equations~\eqref{eq:orthonormalbases} and \eqref{eq:OrthonormalityDDDKKD} imply
orthogonality relations for projectors with $\E$ on the same side (both right or both left):\!
\begin{subequations}
	\label{eq:PexRelations}
\begin{flalign}
\label{eq:PexRelations-a}
&  
\Pc^{\xs \es}_\ell \Pc^{\xsb\es}_\ellb  = 
\delta^{\ell< \ellb}  \delta_{\xs \ks} \Pc^{\xsb \es}_\ellb  
\!+\! \delta^{\ell\ellb}  \delta_{\xs \xsb} \Pc^{\xsb \es}_\ell  
\!+\! \delta^{\ell > \ellb}   \Pc^{\xs \es}_\ell  \delta_{\ks \xsb}
 , \hspace{-5mm} & 
\\ 
\label{eq:PexRelations-b}
&  
\Pc^{\es \xs}_\ell \Pc^{\es \xsb}_\ellb  = 
\delta^{\ell < \ellb}   \Pc^{\es \xs}_\ell  \delta_{\ks \xsb}
\!+\! \delta^{\ell\ellb}  \delta_{\xs \xsb} \Pc^{\es \xsb }_\ell  
\!+\! \delta^{\ell> \ellb}  \delta_{\xs \ks} \Pc^{\es \xsb}_\ellb  
 . \hspace{-5mm} & 
\end{flalign}
\end{subequations}%
The $\delta$ symbols indicate that the first, second, and third terms contribute only for $\ell < \ellb$, $\ell=\ellb$, and $\ell> \ellb$,
respectively. Thus, same-site projectors are orthonormal; different-site products 
with $\E$s on the same side, of the 
type $\Pc^{\xs \es}_\ell \Pc^{\xsb\es}_\ellb$ (or $\Pc^{\es \xs}_\ell \Pc^{\es\xsb}_\ellb$), vanish if the earlier~(later) site hosts a $\D$; 
if it hosts a $\K$, they yield the projector from the other site.
We depict two cases of \Eq{eq:PexRelations-a} below:
\begin{flalign}
\nonumber
& \Pde_\ell \Pde_\ell\!=\!
\raisebox{-0.44\height}{\includegraphics[width=0.683\linewidth]{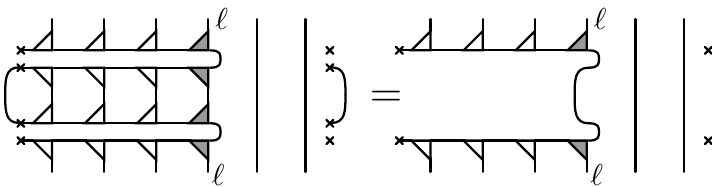}} 
\!\! = \Pde_\ell   \! .
\hspace{-1cm} &
\\  \nonumber
& \Pke_\ell \Pde_\ellb\!=\!
\raisebox{-0.44\height}{\includegraphics[width=0.683\linewidth]{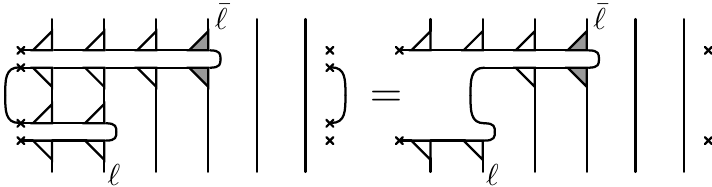}} 
\!\! = \Pde_\ellb   \! .
\hspace{-1cm} &
\end{flalign}

Equation \eqref{eq:PexRelations-a} was first written down in that form 
in Ref.~\cite{Lee2021}, Eq.~(29), in the context of NRG. There, one deals
exclusively with left-normalized states, and sites
to the right of the orthogonality center are treated purely
as environmental degrees of freedom, described by product states.
Equation \eqref{eq:PexRelations-b} is the counterpart of
\eqref{eq:PexRelations-a} for right-normalized states. 

Projector products with $\E$s in the middle, $\Pc^{\xs\es}_{\ell} \Pc^{\es\xsb}_{\ellb}$,  and $\ell < \ellb$, again yield projectors. We denote them 
by 
\begin{align}
\Pxxb_{\ell\ellb} &= \Pc^{\xs\es}_{\ell} \Pc^{\es\xsb}_{\ellb}  \quad 
(0 \le \ell < \ellb \le \eLL\!+\!1) \, ,\hspace{-3cm}
\\
\nonumber \Pkk_{\ell\ellb} & = 
\raisebox{-0.45\height}{\includegraphics[width=0.317\linewidth]{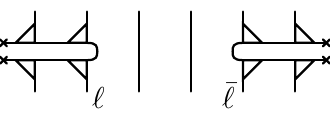}}  \, , 
& 
\Pkd_{\ell\ellb} & = 
\raisebox{-0.45\height}{\includegraphics[width=0.317\linewidth]{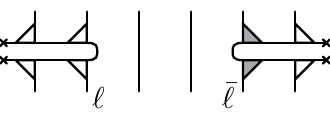}} 
 \, , 
\\[-1.5mm]
\nonumber \Pdk_{\ell\ellb} & = 
\raisebox{-0.45\height}{\includegraphics[width=0.317\linewidth]{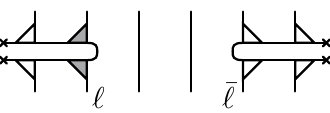}} 
 \,  , 
& 
\Pdd_{\ell\ellb} & = 
\raisebox{-0.45\height}{\includegraphics[width=0.317\linewidth]{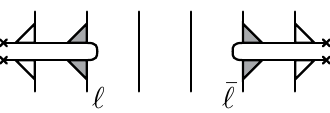}} 
 \, . 
\end{align}
They have local unit operators
on $n=\ellb-(\ell+1)$ contiguous sites, sandwiched between any combination of 
$\K$ and $\D$ projectors to the left and right. In this sense, 
they generalize Eqs.~\eqref{eq:0-1-two-site-projectors} and will be called
generalized \textit{local} \nsite\ projectors.
They fulfill numerous orthogonality relations following directly from Eqs.~\eqref{eq:PexRelations}. For example: 
\begin{subequations}
\label{eq:PxxnOrthonormality}
\begin{flalign}
\label{eq:PxxnOrthonormality-a}
& & \hspace{-7mm} \Pc^{\xs \xsb}_{\ell\ellb} \Pc^{\xsp \xsbp}_{\ell\ellb} & =
\delta^{\xs \xsp} \delta^{\xsb \, \xsbp} \Pc^{\xs \xsb}_{\ell\ellb} 
, \hspace{-9cm}  & \hspace{-9cm}  & & 
\\ 
\label{eq:PxxnOrthonormality-b}
& & \hspace{-8mm} \forall  \ell \!<\! \ellp\!: \;
\Pc^{\ds \xsb}_{\ell\ellb} \Pc^{\xsp \, \xsbp}_{\ellp \ellb{}'} &= 0 \, ,
\hspace{-0.75cm}
& 
\forall  \ellb \!<\! \ellbp \!\!:
\Pc^{\xs \xsb}_{\ell \ellb} \Pc^{\xsp \ds }_{\ellp  \ellbp} & = 0 \, ,
\hspace{-1cm} &
\\ 
\label{eq:PxxnOrthonormality-c}
& & \hspace{-8mm} \Pc^{\ds \xsb}_{\ell\ellb} \Pc^{\ds \xsbp}_{\ellp \ellbp} &\sim 
\delta_{\ell \ellp} 
\, , \hspace{-0.75cm}
& 
\Pc^{\xs \ds}_{\ell\ellb} \Pc^{\xsp \ds}_{\ellp \ellbp} & \sim 
\delta_{\ellb \ellbp} . 
\hspace{-1cm} & 
\end{flalign}
\end{subequations}
Thus, two projectors having the same site indices are orthonormal;
projector products involving a $\D$ on a site earlier or later than all other indexed sites vanish;
those involving two $\D$s on the same side but different sites vanish, too. Some of these relations are illustrated below:
\begin{flalign}
\nonumber
& 
\Pc^{\ds \ks}_{\ell\ellb} \Pc^{\ds \ks}_{\ell\ellb} =  
\raisebox{-0.45\height}{\includegraphics[width=0.7\linewidth]{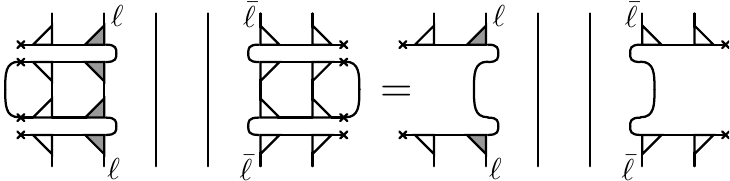}} 
\!\!\!\!\!\! = \Pdk_{\ell \ellb}   \! ,
\hspace{-1cm} &
\\
\nonumber
& 
\Pc^{\ds \ks}_{\ell\ellb} \Pc^{\ks \ks}_{\ellp \ellbp} = 
\raisebox{-0.45\height}{\includegraphics[width=0.317\linewidth]{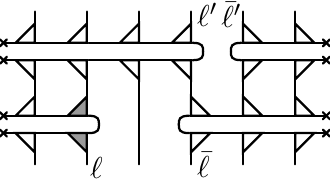}}  
\;=\,  0
 \, , 
 \hspace{-1cm} &
\\
\nonumber
& 
\Pc^{\ds \ks}_{\ell\ellb} \Pc^{\ds \ds}_{\ell \ellbp} = 
\raisebox{-0.425\height}{\includegraphics[width=0.683\linewidth]{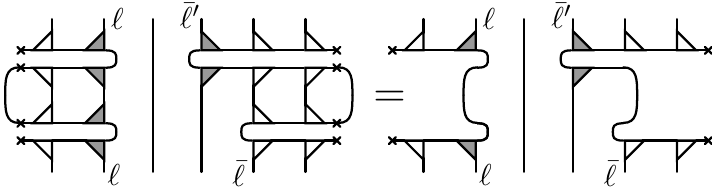}} 
\!\!\!\!\!\! = \Pdd_{\ell \ellbp}   .
\hspace{-1cm} &
\end{flalign}

\Eq{eq:PQ_recursion} implies another useful property (for $\ellb\!-\!\ell\!>\!1$),
\begin{align}
\label{eq:DecomposePKXorXK}
	\Pc^{\ks \xsb}_{\ell \ellb} & =
	\Pc^{\ks \xsb}_{\ellplusone, \ellb}  + \Pc^{\ds \xsb}_{\ellplusone \ellb} \, , 
	\quad
	\Pc^{\xs \ks}_{\ell \ellb} = 
	\Pc^{\xs \ks}_{\ell, \ellb-1}  + \Pc^{\xs \ds}_{\ell, \ellb-1} \, ,
\end{align}
reflecting \Eq{eq:TwoSiteCompleteness}. Thus, a $\K$ on a given site 
$\ell$ (or $\ellb)$ can be decomposed into $\K$ and $\D$
on the inner neighboring site $\ell+1$ (or $\ellb-1$), thereby 
expressing one projector through two that both target
one less site. This decomposition will be used repeatedly below.

\subsection{Local $n$-site projectors, 
$\Pnsite_\ell$}
\label{sec:GeneralizedBondProjectors}

The $\K\K$ projectors merit special attention. For $\ellb - \ell=1$, $2$ or $3$, they correspond to
the bond, \onesite\ and \twosite\ projectors introduced
in Eqs.~\eqref{eq:0-1-two-site-projectors}.  These can be expressed as
\begin{align}
\label{eq:0s1s2s}
 \Pc_\ell^\mbond = \Pkk_{\ell, \ellplusone} , 
\quad \Ponesite_\ell = \Pkk_{\ellminusone,\ellplusone} , 
\quad
\Pc_\ell^\mtwosite  = \Pkk_{\ellminusone,\ell+2} \, .
\end{align}
Generalizing the notation of \eqref{eq:0s1s2s}, we define 
a set of local \nsite\ projectors (for $n\ge 0$ and $\ell \in [1,\eLL\!+\!1\!-\!n]$) as:
\begin{align}
\Pc^{\mnsite}_{\ell} &= \Pkk_{\ell-1,\ell+n} \, = \, 
\raisebox{-0.35\height}{\includegraphics[width=0.317\linewidth]{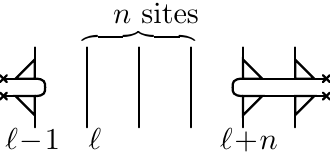}} 
\; .
\end{align}
Then $\Pzerosite_\ell = \Pbond_{\ellminusone}$, and for 
$n\ge 1$, these projectors span the spaces of variations of $\ket{\Psi}$ on $n$ contiguous sites from $\ell$ to $\ell+n-1$. However,
projectors $\Pnsite_{\ell}$ and $\Pnsite_{\ell'}$ with $\ell \neq \ellp$ are not orthogonal. Instead, the following relations hold for all $\ell < \ellp$,
\begin{flalign}
		\label{eq:PNellellp}
\Pnsite_\ell \Pnsite_\ellp  & = \Pnminusonesite_\ellplusone \Pnsite_\ellp
		 = \Pnsite_\ell \Pnminusonesite_\ellp  
  = \Pnminusonesite_\ellplusone \Pnminusonesite_\ellp
		  \, , \hspace{-0.5cm} &  
\end{flalign}
as can be verified using \Eqs{eq:PexRelations}. 
For example, for
\begin{align} 
\nonumber 
\Pnsite_\ell \Pnsite_\ellp
& =
\raisebox{-0.45\height}{\includegraphics[width=0.716\linewidth]{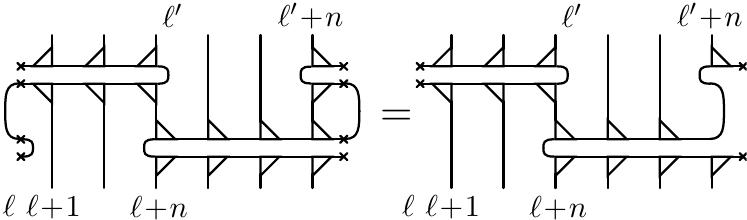}}  ,
\\ 
 \nonumber
\Pnminusonesite_\ellplusone \Pnsite_\ellp
& =
\raisebox{-0.45\height}{\includegraphics[width=0.716\linewidth]{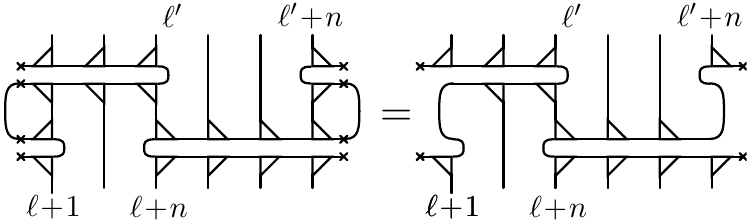}}  ,
\end{align}
we obtain the same result in both cases. In particular, for $n \ge 1$, two \nsite\ projectors mismatched by one site
yield an $(n\!-\!1)$-site projector, 
\begin{align}
\label{eq:PnellPnellplusone}
\Pc^\mnsite_{\ell}\Pc^\mnsite_{\ell+1}  = \Pc^\mnminusonesite_{\ell+1} 
\hspace{1.9cm} \\ 
\nonumber 
\raisebox{-0.45\height}{\includegraphics[width=0.75\linewidth]{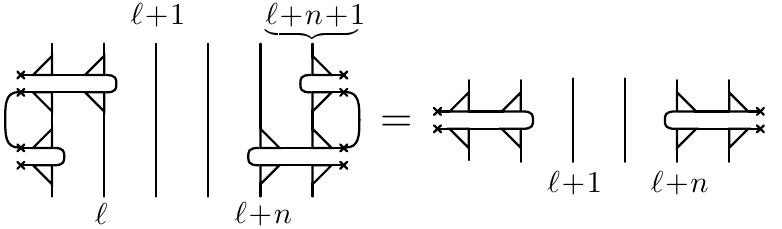}} . 
\end{align}

Orthogonalized versions of the $\Pnsite_\ell$ projectors 
will be constructed in the next subsection. Here, we collect
some properties, following from Eq.~\eqref{eq:PexRelations},
that will be needed for that purpose:
\begin{subequations}
\label{subeq:PnellProperties}
\begin{align}
\label{eq:EarlyLeftDLateRightDYieldZeroPs1}
\forall \ell < \ellp\!\!: \; & & 
\Pc^{\ds \xsb}_{\ell\ellb} \Pc^{\mnsite}_{\ellp} 
& = 0  \, , 
\\
\forall (\ell\!+\!n) \!\le \! \ellbp\!\!: \;  & & 
\Pc^{\mnsite}_\ell \Pc^{\xsp \ds}_{\ellp\ellbp} & = 0 \, . 
\end{align}
\end{subequations}
Thus, $\Pnsite_\ell$ is annihilated by a left $\D$ 
on its left or a right $\D$ 
on its right. For example,
\begin{align}
	  \nonumber
\Pnsite_\ell \Pkd_{\ellp \ellbp}
& =
\raisebox{-0.45\height}{\includegraphics[width=0.35\linewidth]{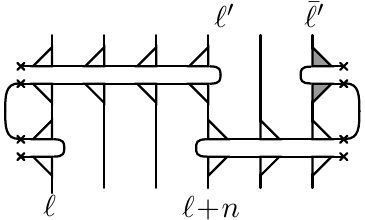}}  = 0 \, . 
\end{align}

Using  \Eq{eq:DecomposePKXorXK}, $\Pnsite_\ell$ can be expressed through two $(n\!-\!1)$s projectors:
\begin{align}
\nonumber 
\Pnsite_{\ell}
&  =  \Pkk_{\ell,\ell+n}+ \Pdk_{\ell,\ell+n}  
=  \Pkk_{\ell-1,\ell+n-1} + \Pkd_{\ell-1,\ell+n-1} 
\\ 
\label{eq:Pn-decompose}
&  = \Pc^\mnminusonesite_{\ell+1} + \Pdk_{\ell,\ell+n}  
= \Pc^\mnminusonesite_{\ell} + \Pkd_{\ell-1,\ell+n-1} 
\\ \nonumber
& = \, \raisebox{-0.48\height}{\includegraphics[width=0.733\linewidth]{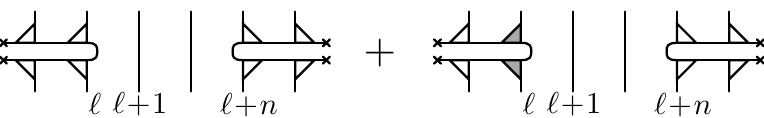}} 
\\ \nonumber
& = \, \raisebox{-0.55\height}{\includegraphics[width=0.733\linewidth]{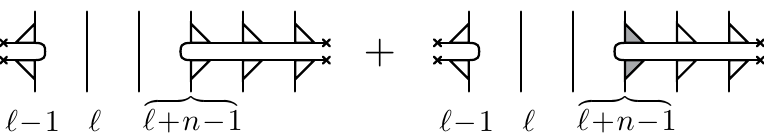}} \,  . 
\end{align}
The existence of two different decompositions of $\Pc^{\mnsite}_{\ell}$,
mimicking \Eq{eq:OneSiteCompletness}, 
reflects the gauge freedom of MPSs. This
can be exploited to write $\Pdk_{\ell,\ell+n}$ as $\Pc^\mnminusonesite_{\ell} 
+ \Pkd_{\ell-1,\ell-1+n} - \Pc^\mnminusonesite_{\ell+1}$,
converting $\D\K$ to $\K\D$, or vice versa. Repeated use yields an identity
that will be useful below:
\begin{flalign}
	\label{eq:KD-to-DK-conversion}
	& \sum_{\ell=\ellb}^\ellp \Pdk_{\ell,\ell+n}   
	 = \Pc^\mnminusonesite_{\ellb} \! + \sum_{\ell = \ellb}^{\ellp} \Pkd_{\ell-1,\ell-1+n} \! - \! \Pc^\mnminusonesite_{\ellp+1}  . 
	 \hspace{-1cm} &
\end{flalign}

\subsection{Global \nsite\ projectors, $\Pnsite$}
\label{sec:ProjectorsnSiteVariations}
\label{sec:Id_nsite_decomposition}

We now are ready to define the \nsite\ spaces $\doubleVnsite$.
For $n =0$, we define $\doubleVzerosite = \mathrm{span}\{\ket{\Psi}\}$. 
For $n\ge 1$, we define $\doubleVnsite$
as the span of $\ket{\Psi}$ and all states 
$\ket{\Psi'}$ differing from it on at most $n$ contiguous sites: 
\begin{flalign}
\label{eq:DefineNsiteTangentSpaceInformal}
 \doubleVnsite & = 
\mathrm{span} \Bigl\{ \!
\raisebox{-0.4\height}{
 \includegraphics[width=0.317\linewidth]{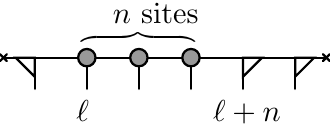}} \; | \; \ell \in [1, \eLL\!+\!1\!-\! n] \Bigr\}  . 
 \hspace{-1cm} & 
\end{flalign}
For $n\!=\!1$, $\doubleV^\monesite$ is the tangent space of $\ket{\Psi}$. 
More concretely, $\doubleV^\mnsite$ is defined as the image of all 
local \nsite\ projectors:
\begin{flalign}
& \doubleV^\mnsite = 
\mathrm{span}\Bigl\{\im(\Pc^\mnsite_1), \im(\Pc^\mnsite_2), \dots, 
\im(\Pc^\mnsite_{\seLL+1-n})\Bigr\}  \, . 
\hspace{-1cm} &
\end{flalign}
For any $n'\leq n$, the image $\im(\Pc^{n'\mr{s}}_{\ell})$ is 
by construction fully contained in the image 
$\im(\Pc^\mnsite_{\ell})$, hence $\doubleV^{n'\mr{s}}$ is a subspace of  $\doubleV^\mnsite$, implying the nested hierarchy \eqref{eq:NestingVnsite}. 

Let $\Pnsite$ be the  projector having $\doubleV^\mnsite$ as image;
then, $\rm{im}(\Pnsite)$ contains $\rm{im}(\Pnsite_\ell)$
for all $\ell \in [1,\eLL+1-n]$. Formally, $\Pnsite$ has the 
defining properties
\begin{subequations}
\begin{align}
	\label{eq:DefineNTangentSpaceProjectorEarly}
	\bigl(\Pnsite\bigr)^2 &= \Pnsite \, , \qquad 
	\Pnsite \Pnsite_\ell = \Pnsite_\ell \, ,
	\\
	\label{eq:DefineNTangentSpaceProjectorEarly_2}
	\Pnsite_\ell \ket{\Phi} &= 0 \; \forall \ell
	\implies
	\Pnsite \ket{\Phi} = 0 \, . 
\end{align}
\end{subequations}
Moreover, the nested structure of the $\doubleVnsite$s implies
\begin{align}
\label{eq:PnPnprimenesting}
\forall n' < n : \quad	\Pnsite \Pnpsite = \Pnpsite \, . 
\end{align}%
 
Let us construct $\Pnsite$ explicitly. 
Simply summing up the local projectors $\Pc^\mnsite_{\ell}$ does not yield a projector because the images of $\Pc^\mnsite_{\ell}$ and $\Pc^\mnsite_{\ell'}$ are not orthogonal. A set of mutually orthogonal local projectors
can be obtained by 
projecting out the overlap between $\Pc^\mnsite_{\ell}$ and $\Pc^\mnsite_{\ell\pm1}$. 
We thus define 
\begin{align}
\Pnsite_{\ell\lessgtr} = 
\Pnsite_{\ell} \bigl(\doubleI[\doubleV] - \Pnsite_{\ell \pm 1}
\bigr) \, , 
\end{align}
so that $\Pnsite_{\ell \lessgtr} \Pnsite_\ellp = 0$
holds for neighboring $\ell, \ellp$ with $\ell \lessgtr \ellp$. 
It suffices to orthogonalize \nsite\ projectors 
mismatched by \textit{one} site, since from these
we can select a set of projectors mutually orthogonal
on all sites. Indeed, 
\Eqs{eq:PnellPnellplusone} and \eqref{eq:Pn-decompose} 
 yield $(n\!-\!1$)-site projectors containing $\D$s, 
\begin{subequations}
\label{subeq:Pns_orthogonalization}
\begin{align}
\label{eq:Pns_orthogonalization-a}
\Pc^\mnsite_{\ell <} 
& = \Pc^\mnsite_{\ell} - \Pc^\mnminusonesite_\ellplusone  = 
\Pdk_{\ell,\ell+n} \, , 
\\ 
\label{eq:Pns_orthogonalization-b}
 \Pc^\mnsite_{\ell > } 
& = \Pc^\mnsite_{\ell} - \Pc^\mnminusonesite_\ell  = 
\Pkd_{\ellminusone,\ellminusone+n} \, , 
\end{align}
\end{subequations}
and the $\D$s ensure the orthonormality relations
(cf.\ \eqref{eq:PxxnOrthonormality})

\vspace{-1.5\baselineskip}
\begin{subequations}\label{subeq:orthogonalityPDKKD}
\begin{align}
& & \Pnsite_{\ell \lessgtr} \Pnsite_{\ellp \lessgtr}&= \delta_{\ell\ell'} 
\Pnsite_{\ell \lessgtr} \, , 
\\ 
\forall \ell < \ell' \!\!: & & \Pnsite_{\ell < } \Pnsite_{\ellp > }
&= 0 \, , 
\\ 
\forall \ell \lessgtr  \ell' \!\!: & & 
\Pnsite_{\ell \lessgtr} \Pc_{\ell'}^{\mnsite} & = 0 \, . 
\end{align}
\end{subequations}
 
 These equations have a remarkable implication:  
for any choice of $\ellp \in [1, \eLL\!-\!n\!+\!1]$, 
the projectors 
 $\Pnsite_{\ell <}$ for $\ell \in [1, \ellp-1]$, 
 $\Pnsite_\ellp$, and $\Pnsite_{\ell > }$ for $\ell \in [\ellp+1 , \eLL+1-n]$ 
form an orthonormal set, and this set contains a $\Pnsite_\ell$ (in projected form) for \textit{every} $\ell \in [1,\eLL+1-n]$.
We define the global \nsite\ projector as their sum, 
\begin{align}
\label{eq:PnsiteDefinition}
\Pc^{\mnsite} &
=   \sum_{\ell=1}^{\ellp-1}  \Pnsite_{\ell < } \; 
+ \; \Pc^{\mnsite}_{\ellp} 
\; +  \sum_{\ell=\ellp+1}^{\scripteLL+1-n} \Pnsite_{\ell > } \, 
\\ \nonumber
& 
=  \hspace{4mm} \sum_{\ell=1}^{\ellp-1}
 \hspace{2mm} 
\raisebox{-0.5\height}{\includegraphics[width=0.317\linewidth]{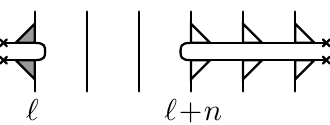}} \; +
 \hspace{1.25mm} 
\raisebox{-0.5\height}{\includegraphics[width=0.317\linewidth]{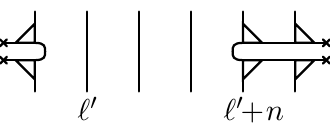}} 
\\ \nonumber
& \phantom{=} + 
\sum_{\ell=\ellp+1}^{\scripteLL+1-n} \hspace{-0.5mm}
\raisebox{-0.5\height}{\includegraphics[width=0.333\linewidth]{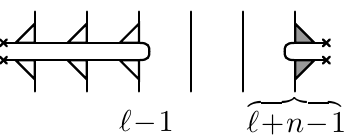}} \, .
\end{align}
Here, $\ellp$ may be chosen freely
as convenience dictates; 
different choices are equivalent, being related by Eqs.~\eqref{eq:Pn-decompose}. The orthogonality relations
\eqref{subeq:orthogonalityPDKKD} 
ensure the properties 
\eqref{eq:DefineNTangentSpaceProjectorEarly}. For example,
\begin{align}
	\Pnsite \Pnsite_\ellp = 0 + \Pnsite_\ellp  \Pnsite_\ellp + 0  
	=  \Pnsite_\ellp \, .  
\end{align}
The property \eqref{eq:DefineNTangentSpaceProjectorEarly_2} is ensured by orthogonalizing
$\Pnsite_{\ell}$ w.r.t. each other and thus never including states with $\Pnsite_{\ell}\ket{\Phi} = 0 \; \forall \ell$. 
This confirms that $\rm{im}(\Pnsite)$ contains
$\rm{im}(\Pnsite_\ell)$ for all $\ell\in [1, \eLL+1-n]$; thus,
$\Pnsite$ indeed is the desired projector having
$\doubleV^\mnsite$ as image. 
Evaluating \Eq{eq:PnsiteDefinition} using the 
middle expressions from \eqref{subeq:Pns_orthogonalization}, we obtain
\begin{subequations}
\label{subeq:Pnsite_explicitAlternative}
\begin{align}
\label{eq:Pnsite_explicitAlternative}
\Pc^{\mnsite} &
=   \sum_{\ell=1}^{\scripteLL+1-n} \Pc^\mnsite_{\ell} 
- \sum_{\ell=1}^{\scripteLL-n} \Pc^\mnminusonesite_{\ell+1} \, 
\\ \nonumber
& \hspace{-5mm} 
= \sum_{\ell=1}^{\scripteLL+1-n}
\raisebox{-0.5\height}{\includegraphics[width=0.317\linewidth]{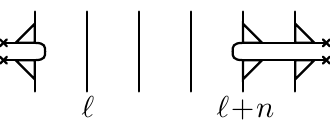}} \; - 
\sum_{\ell=1}^{\scripteLL-n} \hspace{1.25mm} 
\raisebox{-0.5\height}{\includegraphics[width=0.317\linewidth]{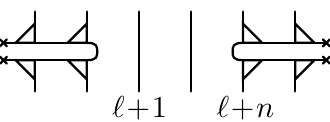}} \, ,
\end{align}
expressing $\Pnsite$ through local \nsite\ and 
$(n-1)$s projectors in a manner manifestly 
independent of $\ellp$, and not involving an $\D$ sectors.
 The occurrence of the first term, a sum over all $\Pnsite_\ell$, is no surprise; the nontrivial part of the above construction was establishing the form of the second term, needed to ensure that
$\Pnsite$ is a projector. 
Note that \Eq{eq:Pnsite_explicitAlternative} directly 
implies
property \eqref{eq:DefineNTangentSpaceProjectorEarly_2}.
Alternatively, we can use the rightmost forms of 
\eqref{subeq:Pns_orthogonalization}
in \eqref{eq:PnsiteDefinition} to obtain  
\begin{align}
\Pnsite  \label{eq:Pnsite_explicit}
 &= \sum_{\ell=1}^{\ellp} \Pc_{\ell,\ell+n}^{\ds \ks} + 
 \Pc^{\ks \ks}_{\ellp,\ellp+n} 
+ \sum_{\ell=\ellp}^{\scripteLL-n} \Pc_{\ell,\ell+n}^{\ks \ds}
\, ,
\end{align}
\end{subequations}
now expressed purely through $(n\!-\!1)$s projectors,
with all but one involving $\D$ sectors.

For $n=1$, \Eqs{subeq:Pnsite_explicitAlternative} reproduce
the well-known tangent space projector, 
\begin{subequations}
\label{subeq:TangentExplicit}
\begin{flalign}
\label{eq:P1site_explicitAlternativeTangent}
\Ponesite &
=   \sum_{\ell=1}^{\scripteLL} \Ponesite_{\ell} 
- \sum_{\ell=1}^{\scripteLL-1} \Pbond_\ell \,  & 
\\ \nonumber
& = \sum_{\ell = 1}^{\scripteLL}
\! \raisebox{-5mm}{
 \includegraphics[width=0.317\linewidth]{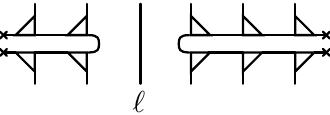}} 
\;  - \sum_{\ell = 1}^{\scripteLL-1} 
\! \raisebox{-5mm}{
 \includegraphics[width=0.317\linewidth]{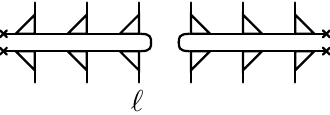}} \; ,
  \hspace{-1cm} & \nonumber
  \\ 
  \label{eq:P1site_explicit}
 &= \sum_{\ell=1}^{\ellp} \Pc_{\ell,\ell+1}^{\ds \ks} + 
 \Pc^{\ks \ks}_{\ellp,\ellp+1} 
+ \sum_{\ell=\ellp}^{\scripteLL-1} \Pc_{\ell,\ell+1}^{\ks \ds}
\, 
\\ \nonumber
&= 
\sum_{\ell = 1}^{\ellp}
\! \raisebox{-5mm}{
 \includegraphics[width=0.317\linewidth]{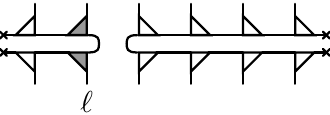}} 
\;  + 
\! \raisebox{-5mm}{
 \includegraphics[width=0.317\linewidth]{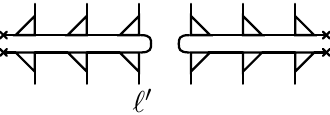}} 
  \hspace{-1cm} & \nonumber 
  \\ & \phantom{=} + 
  \sum_{\ell = \ellp}^{\scripteLL -1 }
\! \raisebox{-5mm}{
 \includegraphics[width=0.317\linewidth]{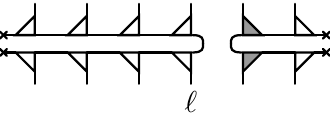}} \, . 
 \nonumber
\end{flalign} 
\end{subequations}%
These expressions are widely used in MPS algorithms based on tangent space concepts, such as time evolution using the time-dependent variational principle (TDVP) \cite{Lubich2015a,Haegeman2016,ZaunerStauber2018,Vanderstraeten2019,Li2022}. 
The form \eqref{eq:P1site_explicitAlternativeTangent}, or 
\eqref{eq:P1site_explicit} with the choice $\ellp=\eLL-1$,  was first given 
Lubich, Oseledts and Vandereycken \cite{Lubich2015a} (Theorem 3.1), and transcribed into MPS notation in Ref.~\cite{Haegeman2016}. 
In these works, it was derived in a different manner than here,
using arguments invoking gauge invariance. Our derivation has the
advantage that it generalizes directly to \nsite\ projectors.
For $n=2$, our expression \eqref{eq:Pnsite_explicitAlternative} 
for $\Ptwosite$ reproduces the projector proposed
in Ref.~\cite{Haegeman2016} for \twosite\ TDVP:
\begin{align}
	\label{eq:P2siteDiagrams}
	\Ptwosite = \!
	\sum_{\ell=1}^{\scripteLL-1}\!
\raisebox{-5.3mm}{
\includegraphics[width=0.317\linewidth]{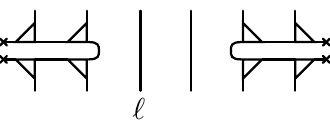}}\;
	- \!\sum_{\ell=2}^{\scripteLL-1}\!
	\raisebox{-5.3mm}{
  \includegraphics[width=0.317\linewidth]{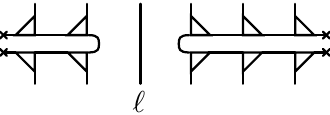}}  \, .
	\end{align}

\subsection{Irreducible global \nsite\ projectors, $\Pc^{n\perp}$}
\label{sec:OrthogonalGlobalnsProjectors}

Our final step is to orthogonalize the global
$\Pnsite$ projectors to obtain mutually \textit{orthogonal} global
\nsite\ projectors, $\Pn$. This step is inspired by 
the observation, made in Ref.~\onlinecite{Hubig2018}, that a given  MPS $\ket{\Psi}$ induces a decomposition of the  full Hilbert space
into mutually orthogonal subspaces, 
\begin{align}
\label{eq:VDecomposition}
\doubleV = \oplus_{n=0}^\seLL \doubleV^{n\perp},
\end{align}
where  $\doubleV^{0\perp}$ is spanned by 
$\ket{\Psi}$, and for $n \ge 1$ each $\doubleV^{n\perp}$ is the complement of $\doubleV^\mnminusonesite$ in $\doubleV^{\mnsite} = \doubleV^\mnminusonesite 
\jvd{\oplus} 
\doubleV^{n\perp}$. 
Each $\doubleV^{n\perp}$ is  
\textit{irreducible}, comprising variations of $\ket{\Psi}$ 
defined on $n$ contiguous sites that are not expressible through variations on subsets of $n' < n$ sites.

The decomposition \eqref{eq:VDecomposition} induces 
a  decomposition of the identity on $\doubleV$
into a sum of irreducible, mutually orthogonal projectors, 
$\Pn$, each with a $\doubleV^{n\perp}$ as image:
\begin{align}
\doubleI[\doubleV] = 
\doubleI[d]^{\otimes \, \scripteLL} 
& = \sum_{n=0}^{\scripteLL} \Pn \, ,
\quad
\label{eq:ProjectorOrthonormality(n)}
	\Pn \Pnp  =  \delta^{nn'} \Pn \, .
\end{align}

We now construct the $\Pn$ projectors  through a Gram-Schmidt procedure.
For $n \ge 1$, we define $\Pn$ by projecting out $\Pc^\mnminusonesite$ from $\Pnsite$, using \Eq{eq:PnPnprimenesting}:
\begin{align}
\label{eq:DefinitionPnperp}
\Pc^{n \perp} & =
\Pnsite \bigl( \doubleI[\doubleV] - \Pc^\mnminusonesite \bigr)
=
\Pnsite - \Pc^\mnminusonesite \, . 
\end{align}
This scheme is initialized by the definition
\begin{subequations}
\label{eq:defineP0}
\begin{flalign}
\label{eq:defineP0-a}
\quad \Pzero  & = \Pzerosite =
\ket{\Psi}\bra{\Psi} && \\ 
\label{eq:P0left}
& =  \Pzerosite_{\seLL+1}  = \Pkk_{\seLL,\seLL+1} 
   && \hspace{-1.3cm} = \;  \raisebox{-0.45\height}{\includegraphics[width=0.317\linewidth]{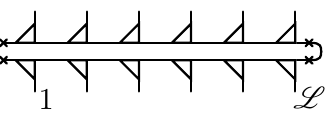}}  \, , \hspace{-1cm} &
\\ 
\label{eq:P0right}
& =  \Pzerosite_{1}  = \Pkk_{0,1} 
&& \hspace{-1.3cm}  =  \; \raisebox{-0.45\height}{\includegraphics[width=0.317\linewidth]{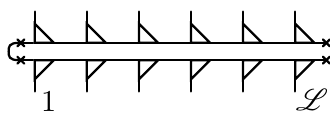}}  \,\, . \hspace{-1cm} &
\end{flalign}
\end{subequations}
The two equivalent forms for $\Pzero$, \eqref{eq:P0left} and \eqref{eq:P0right}, reflect MPS gauge invariance.

For $n=1$, \Eqs{eq:DefinitionPnperp} and
\eqref{eq:Pnsite_explicitAlternative}, 
with $\Pzerosite = \Pzerosite_{1}$, yield
\begin{subequations}
	\label{subeq:P1perp}
\begin{align}
	\label{eq:P1perp-a}
\Pone & = \Ponesite - \Pzerosite
= \sum_{\ell=1}^\scripteLL \Bigl[ \Ponesite_\ell 
- \Pzerosite_\ell \Bigr] 
\\ \nonumber
& = \sum_{\ell=1}^\scripteLL \Bigl[ 
\! \raisebox{-5mm}{
 \includegraphics[width=0.317\linewidth]{Eq/Ponesiteell}} 
 \,  - \,  \raisebox{-5mm}{
 \includegraphics[width=0.317\linewidth]{Eq/Pzerositeell}} 
\Bigr] .
\end{align}
More compact forms are obtained by 
 evaluating \Eq{eq:DefinitionPnperp} using 
\Eq{eq:Pnsite_explicit}, choosing either  $\ellp\!=\!\eLL$ or 0
for $\Pzerosite$:
\begin{align}
\label{eq:P1_left}
\Pone & = \sum_{\ell = 1}^{\scripteLL} \Pdk_{\ell,\ell+1}
 = \sum_{\ell = 1}^{\scripteLL} 
\raisebox{-0.47\height}{\includegraphics[width=0.317\linewidth]{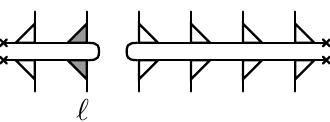}}   
\, , 
\\
 \label{eq:P1_right}
& 
= \sum_{\ell = 1}^{\scripteLL} \Pkd_{\ellminusone,\ell} 
= \sum_{\ell = 1}^{\scripteLL} \;  
  \raisebox{-0.47\height}{\includegraphics[width=0.317\linewidth]{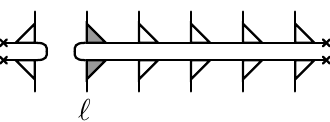}} \, . 
\end{align}
\end{subequations}
Diagrammatically, the latter expressions also
follow directly from 
\eqref{eq:P1perp-a},  using \eqref{eq:OneSiteCompletness}.
That two equivalent forms exist again reflects MPS gauge invariance.

For $n\ge 2$, \Eqs{eq:DefinitionPnperp} and
\eqref{eq:Pnsite_explicitAlternative} yield
\begin{subequations}
\label{subeq:ConstructionPnperp}
\begin{flalign}
\label{eq:ConstructionPnperpAlternative}
\Pc^{n \perp} & 
 =  \sum_{\ell=1}^{\scripteLL+1-n} \! 
 \Bigl[ \Pc^\mnsite_{\ell}  
 - \Pc^\mnminusonesite_{\ellplusone} - \Pc^\mnminusonesite_{\ell}
 + \Pc^{(n-2)\mathrm{s}}_{\ellplusone}
\Bigr] 
\hspace{-1cm} & 
\\ \nonumber
& \hspace{-7mm}
=  \sum_{\ell=1}^{\scripteLL+1-n}
\Bigl[\, 
\raisebox{-0.49\height}{\includegraphics[width=0.317\linewidth]{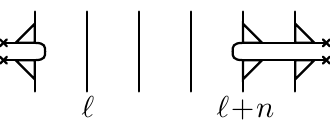}} 
\, - \, \raisebox{-0.49\height}{\includegraphics[width=0.317\linewidth]{Eq/PnellplusoneminusoneKeptKept}} 
\, \Bigr. 
\\ \nonumber
& \hspace{-8mm} 
\hspace{1.375cm}
\Bigl.
- \,\, \raisebox{-0.56\height}{\includegraphics[width=0.317\linewidth]{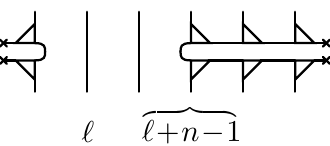}} 
\, + \, 
\raisebox{-0.56\height}{\includegraphics[width=0.317\linewidth]{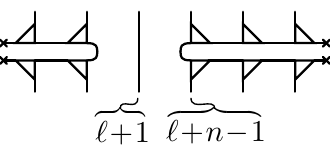}} \Bigr] \, . 
\end{flalign}
A more compact form is obtained 
by evaluating \Eq{eq:DefinitionPnperp} using 
\Eq{eq:Pnsite_explicit},  choosing
$\ellp = \eLL+1-n$ for \textit{both} terms: 
\begin{align}
\nonumber 
\Pn
&= \sum_{\ell=1}^{\scripteLL+1-n} \Pc_{\ell,\ell+n}^{\ds \ks} 
+ \Pc^{\ks\ks}_{\seLL+1-n,\,\seLL+1} 
\\ \nonumber
& \phantom{=} -\sum_{\ell=1}^{\scripteLL+1-n}
\Pc_{\ell,\ell+n-1}^{\ds \ks}
- \Pc^{\ks\ks}_{\seLL+1-n,\,\seLL} - \Pc^{\ks\ds}_{\seLL+1-n,\,\seLL} 
\\
\label{eq:ConstructionPnperp-c}
&= \sum_{\ell=1}^{\scripteLL-(n-1)} \Pc_{\ell,\ell+n-1}^{\ds \ds} 
= \sum_{\ell=1}^{\scripteLL-(n-1)} \hspace{-2mm}
 \!\!\! \raisebox{-6.7mm}{
 \includegraphics[width=0.317\linewidth]{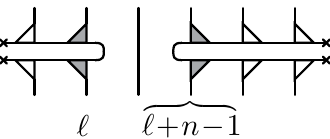}} 
 \, .
\end{align}
\end{subequations}
We used the first and second 
relations in \Eq{eq:DecomposePKXorXK} to
combine the $\sum_\ell$ sums and cancel the remaining
terms. Diagrammatically, \Eq{eq:ConstructionPnperp-c} also
follows directly from 
\eqref{eq:ConstructionPnperpAlternative},  using a relation analogous
to \eqref{eq:DDTwoSiteCompleteness} (with $n-2$ additional 
unit operator lines in the middle). Its form is very natural: 
$n-2$ unit operators are sandwiched between two $\D$s, 
which project out contributions contained in 
$n'$-site projectors with $n'< n$. For future reference
we also display the $n=2$ projector:
\begin{align}
\label{eq:P2_Psi}
	\Ptwo  = \sum_{\ell=1}^{\scripteLL-1} \Pdd_{\ell,\ellplusone}
  = \sum_{\ell=1}^{\scripteLL-1}
 \!\! \raisebox{-4.5mm}{
 \includegraphics[width=0.317\linewidth]{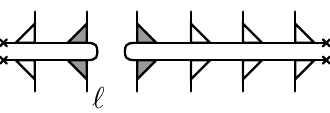}} \, . 
\end{align}
This projector is implicitly used in Ref.~\cite{Hubig2018} 
to compute the \twosite\ variance, as will be recapitulated below.
It also plays a key role in controlled bond expansion algorithms recently developed by us
for performing DMRG ground state searches \cite{Gleis2022} and
TDVP time evolution \cite{Li2022} with \twosite\ accuracy at \onesite\ costs.

Equations~\eqref{eq:defineP0} to 
\eqref{eq:P2_Psi}, giving explicit formulas for $\Pn$ for all $n$, are the main results of the last three sections.

The orthonormality of the $\Pn$, guaranteed by construction, relies on gauge invariance. This is seen when 
verifying orthonormality explicitly. For example,   $\Pone \Pzero = 0$ can be shown in two ways, using either $\Pdk_{\ell,\ellplusone} \Pkk_{\seLL, \, \seLL+1} = 0$  or  $\Pkd_{\ellminusone,\ell} \Pkk_{0,1}  = 0$ (both
 relations hold $\forall \ell \in [1,\eLL]$). 

We continue with some remarks providing intuition about 
the structure of states in the image of $\Pn$. 
The basis states for the spaces $\doubleV^{n\perp}$ can be chosen such that they
involve wavefunctions of the following forms:
\begin{subequations}
\label{subeq:RepresentativeStatesV(n)}	
\begin{flalign}
\label{eq:RepresentativeStatesV(0)}	
& \doubleV^{0\perp}\! : & & \hspace{-10mm}
\raisebox{-0.5\height}{
 \includegraphics[width=0.317\linewidth]{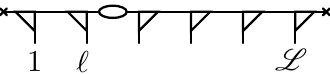}} 
  \quad (\ell \!\in\!  [0,\eLL]),
\hspace{-1cm} & 
\\ 
\label{eq:RepresentativeStatesV(1)}	
& \doubleV^{1\perp}\! : & & \hspace{-10mm}
\raisebox{-0.5\height}{
 \includegraphics[width=0.317\linewidth]{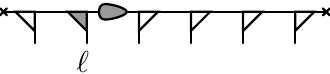}}
 \quad (\ell \!\in\!  [1,\eLL]),
\hspace{-1cm} & 
\\ 
\label{eq:RepresentativeStatesV(1)'}	
& \hphantom{\doubleV^{1\perp}\! :} & & \hspace{-10mm}
  \raisebox{-0.5\height}{
  \includegraphics[width=0.317\linewidth]{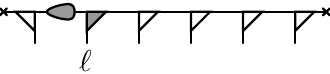}}  
  \quad (\ell \!\in\!  [1,\eLL]),
\hspace{-1cm} & 
\\ 
\label{eq:RepresentativeStatesV(2)}	
& \doubleV^{2\perp}\! : & & \hspace{-10mm}
\raisebox{-0.5\height}{
 \includegraphics[width=0.317\linewidth]{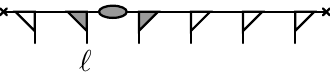}} 
 \quad (\ell \!\in\!  [1,\eLL\!-\!1]),
\hspace{-1cm} & 
\\ 
\label{eq:RepresentativeStatesV(n)}	
& \doubleV^{(n>2)\perp}\! : & & \hspace{-10mm}
\raisebox{-0.5\height}{
 \includegraphics[width=0.317\linewidth]{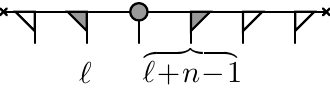}}  
 \quad (\ell \!\in\!  [1,\eLL\!-\!n\! +\!1]).
\hspace{-1cm} & 
\end{flalign}
\end{subequations}
Due to MPS gauge invariance, any choice of $\ell$ 
in \Eq{eq:RepresentativeStatesV(0)}
for $\doubleV^{0\perp}$ yields the \textit{same} wavefunction  $\Psi$.
Gauge invariance also implies that the wavefunctions in \Eqs{eq:RepresentativeStatesV(1)}
and \eqref{eq:RepresentativeStatesV(1)'} for $\doubleV^{1\perp}$ are not all independent;
nevertheless, both forms are useful.
To explicitly construct a complete basis on $\doubleV^{1\perp}$, we can for instance use the form Eq.~\eqref{eq:RepresentativeStatesV(1)} and construct a complete set of 
mutually orthonormal bond matrices of dimension $\Db{}^{A}_{\ell}\times D_{\ell}$ for every bond $\ell$. ($\Db{}^{A,B}_\ell$ are defined near the beginning of Sec.~\ref{sec:OrthogonalProjectors}.)  Note that we could have as well used the form Eq.~\eqref{eq:RepresentativeStatesV(1)'}. Using this construction, we can also explicitly determine the dimension of $\doubleV^{1\perp}$, $\dim \doubleV^{1\perp} = \sum_{\ell=1}^{\scripteLL} \Db{}^{A}_{\ell} D_{\ell}$. 
In the same way, a complete basis with states of the form Eq.~\eqref{eq:RepresentativeStatesV(2)} for $\doubleV^{2\perp}$ can be constructed by constructing a complete set of mutually orthonormal $\Db{}^{A}_{\ell}\times \Db{}^{B}_{\ell+1}$ bond matrices for every bond $\ell$. 
Thus, we find $\dim \doubleV^{2\perp} = \sum_{\ell=1}^{\scripteLL-1} \Db{}^{A}_{\ell} \Db{}^{B}_{\ell+1}$.
A complete basis for $\doubleV^{(n>2)\perp}$ may be characterized by finding, for every $\ell < \eLL-n+1$, a complete set of mutually orthogonal ($n-2$)-site MPS which connect $\Ab_{\ell}$ and $\Bb_{\ell+n-1}$ in Eq.~\eqref{eq:RepresentativeStatesV(n)}. There are $\Db{}^{A}_{\ell} d^{n-2} \Db{}^{B}_{\ell+n-1}$ such MPSs for every $\ell$, i.e. $\dim \doubleV^{(n>2)\perp} = \sum_{\ell=1}^{\scripteLL-n+1} \Db{}^{A}_{\ell} d^{n-2} \Db{}^{B}_{\ell+n-1}$.
The basis states for $\doubleV^{(n>0)\perp}$ 
differ from the reference state $\ket{\Psi}$ in $\doubleV^{0\perp}$  through the replacement of a kept by a discarded space involving precisely one site for $n=1$, and two adjacent sites for $n=2$. For $n>2$, they differ by two discarded spaces and $n\!-\!2$ contiguous sites sandwiched between them, involving virtual bond spaces orthogonal to those from  $\ket{\Psi}$. Therefore, states from $\doubleV^{n\perp}$ and $\doubleV^{n'\perp}$
are manifestly mutually orthogonal if $n\neq n'$. This can be checked via \Eqs{eq:OrthonormalityDDDKKD}, e.g.\ for $\doubleV^{0\perp}$ and $\doubleV^{1\perp}$:
\begin{flalign}
&  \raisebox{-0.35\height}{\includegraphics[width=0.343\linewidth]{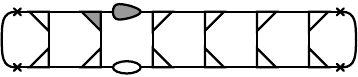}} = 0 \, ,  
  \,   
  \raisebox{-0.35\height}{\includegraphics[width=0.343\linewidth]{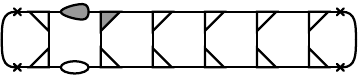}} = 0 \, . \hspace{-1cm} &
\label{eq:ImageOfPKD}
\end{flalign}

States of the form \eqref{subeq:RepresentativeStatesV(n)}	
yield a complete basis for $\doubleV$. This is ensured by our 
Gram-Schmidt construction; but for $\Pone$, the completeness
is not self-evident. For example,
consider a state $\ket{\Psi'}$ of the following form:
\begin{align}
\nonumber
\Psi' &  = 	
\raisebox{-0.4\height}{
 \includegraphics[width=0.317\linewidth]{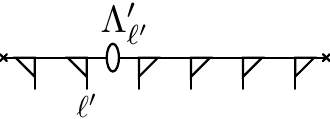}} \, ,
 \;\;
 		\Psi = 	
\raisebox{-0.4\height}{
 \includegraphics[width=0.317\linewidth]{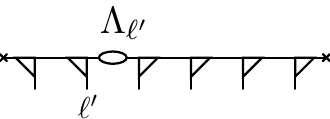}} 
 \, ,  
\\
 \langle \Psi | \Psi' \rangle & =
 \raisebox{-0.3\height}{
 \includegraphics[width=0.566\linewidth]{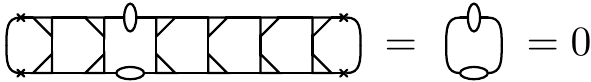}} \, .
\end{align}
It differs from $\ket{\Psi}$ only in the $\K$ space of bond of $\ellp$, 
having a bond matrix $ \Lambda'_\ellp$ orthogonal to the $\Lambda_\ellp$ of $\ket{\Psi}$. Since  $|\Psi'\rangle$ is orthogonal to $\ket{\Psi}$
it does not lie in $\doubleV^{0\perp}$, but it is not immediately apparent
that it lies in $\im(\Pone)$. To see that it does,
we rewrite \Eq{eq:P1_left} such that it 
contains $\D\K$s to the left of site $\ell$ and
$\K\D$s to its right, using  \Eq{eq:KD-to-DK-conversion} 
 (with $\ellb$, $\ellp$ there replaced by $\ellp+1$, $\eLL$):
\begin{align}	
	\label{eq:P(1)Reexpressed-a}
& 	\Pone  = \! 
	\sum_{\ell = 1}^{\ellp-1} \Pdk_{\ell ,\ell+1}  + \Ponesite_\ellp
	+ \!\! \!\sum_{\ell = \ellp+1}^{\scripteLL} \!\! \Pkd_{\ell-1, \ell}
    - \Pkk_{\seLL, \, \seLL+1} \, . 
\end{align}
When evaluating $\Pone \ket{\Psi'}$ using this 
form, and recalling that $\Pkk_{\seLL, \, \seLL+1} = \ket{\Psi}\bra{\Psi}$, we find that
all terms but the second yield zero, and the second yields $\ket{\Psi'}$,
as claimed above.
In this manner, one sees that the image
of  $\Pone$ indeed contains all single-site and single-bond variations of $\ket{\Psi}$ that are orthogonal to $\ket{\Psi}$.

To conclude this section, we remark that the nested structure of $\doubleV$ is an integral part for (thermo)dynamical computations using the NRG~\cite{Peters2006,Weichselbaum2007,Weichselbaum2012a}, although a slightly different structure from $\Pc^{n\perp}$ is used to systematically span the full Hilbert space. 
While the chain considered in NRG is in principle semi-infinite, this chain is in practice cut off naturally by thermal weights~\cite{Weichselbaum2007,Weichselbaum2012a}.
The resulting chain length $\eLL$ increases logarithmically 
with decreasing temperature.
In NRG, the so-called Anders-Schiller basis~\cite{Anders2005} is routinely used, which decomposes the full identity as follows:
\begin{align}
\label{eq:AndersSchiller}
\doubleI[\doubleV] 
= \sum_{\ell=1}^{\scripteLL} \raisebox{-0.45\height}{\includegraphics[width=0.317\linewidth]{Eq/PDiscardedEnv}}
\, .
\end{align}
Here,  all states of the parent space associated with the last site, $\eLL$,
are considered discarded, i.e.\ the kept space of site $\eLL$ has dimension 0.
The projectors occurring in \Eq{eq:AndersSchiller} are constructed from approximate
eigenstates of the Hamiltonian, so that this decomposition of unity can be used, e.g.,\  to explicitly construct time-evolution operators~\cite{Anders2005}, full thermal density matrices~\cite{Weichselbaum2007,Weichselbaum2012a} or evaluate Lehmann representations for two-point~\cite{Weichselbaum2007} or recently even multi-point~\cite{Kugler2021,Lee2021} spectral functions.

\section{Energy variance}
\label{eq:EnergyvarianceTangentProjector}

The decomposition of the identity $\doubleI[\doubleV]$ into 
mutually orthogonal $n$-site projections can be used to 
similarly split the energy variance,  $\variance
= \|(H \! - \! E)\Psi\|^2$, of a state with average energy 
$ E = \bra{\Psi}H\ket{\Psi}$ into $n$-site contributions.
For $n=1$ and $2$, these were given in Ref.~\cite{Hubig2018}.
Here, we extend their analysis to general $n$:
\begin{subequations}
	\label{subeq:EnergyVariance}
\begin{flalign}
\label{eq:EnergyVarianceResult}
\variance &= 
\sum_{n=0}^{\scripteLL} \bra{\Psi} (H\!-\!E) \Pn (H\!-\!E)
\ket{\Psi} 
= \sum_{n=1}^{\scripteLL} \variance^{n\perp} \, , 
\hspace{-1cm} & 
\\ \label{eq:EnergyVariance}
\variance^{n\perp} & =   \| \Pn H \Psi\|^2 
\\ 
& = \begin{cases} 
\sum_{\ell=1}^\scripteLL \| \Pdk_{\ell,\ell+1}  H \Psi \|^2
& (n=1) ,
\\[1.5mm] 
\sum_{\ell=1}^{\scripteLL +1 - n} 
\| \Pdd_{\ell,\ell+n-1} H \Psi \|^2 
& (n\ge 2) \, .
\end{cases} 
 \hspace{-1cm} & 
\end{flalign}
\end{subequations}
In the first line, we used 
\eqref{eq:ProjectorOrthonormality(n)}, $\doubleI[\doubleV]= \sum_{n=0}^\seLL \Pn$; since $\Pzero=\ket{\Psi}\bra{\Psi}$
and $\Pc^{(n>0)\perp} \ket{\Psi}=0$, 
the potentially large contributions linear and quadratic in $E$ 
drop out. 
This convenient feature, emphasized in Ref.~\cite
{Hubig2018}, significantly improves the accuracy of the determination of $\variance$. 
The cumulative \nsite\ variance is defined as 
$\variance^\mnsite 
= \sum_{n'=1}^{n} \variance^{n'\perp}$.

\begin{figure}[b!]
 \includegraphics[width=\linewidth]{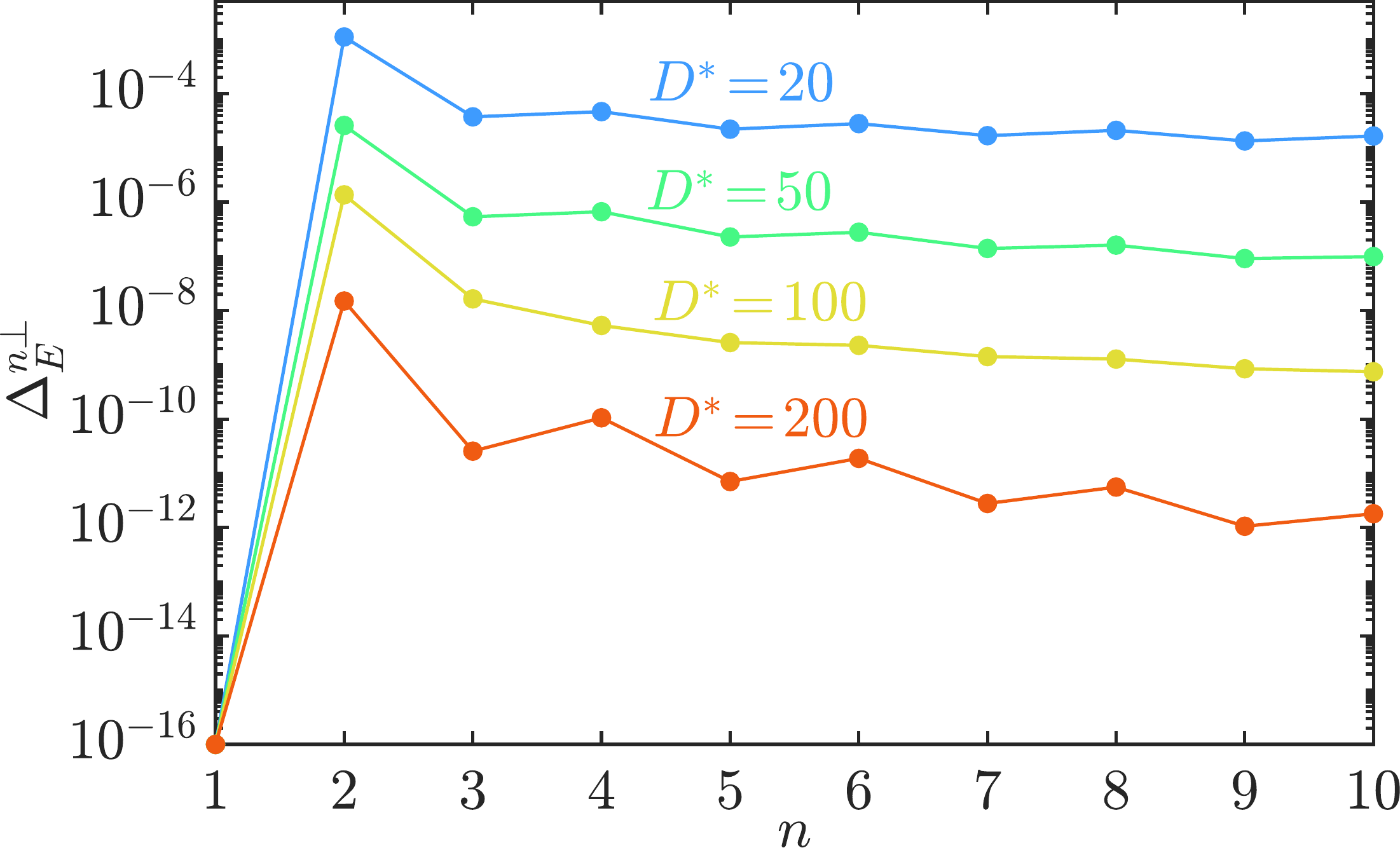} 
  \vspace{-7mm}
 \caption{The $n$-site variance, 
$\variance^{n\perp}$, of the $\eLL=40$ Haldane-Shastry model for different $\Dast$. $\variance^{1\perp}$ can in principle 
always be converged to
numerically zero (i.e.\ $\variance^{1\perp} \lesssim 10^{-16}$) 
by extensive DMRG sweeping; this being the case here, we plot it symbolically at $\variance^{1\perp}=10^{-16}$.
In practice it suffices to sweep until 
$\variance^{1\perp} \ll \variance^{2\perp}$, since 
the variance is dominated by $\variance^{2\perp}$.
 \label{fig:variance_HS}} 
 \end{figure}

Expressed diagrammatically, the \onesite\ and \nsite\ variance 
are 
\begin{subequations}
	\label{eq:VarianceNSitePerpDiagrams}
\begin{flalign}
\label{eq:Variance1SitePerpDiagrams}
	 \variance^{1\perp} & = \sum_{\ell=1}^{\scripteLL}
 \raisebox{-7.5mm}{
 \includegraphics[width=0.166\linewidth]{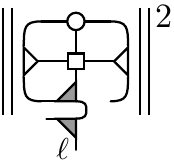}} 
 =
 \sum_{\ell=1}^{\scripteLL}
 \raisebox{-5.3mm}{
 \includegraphics[width=0.166\linewidth]{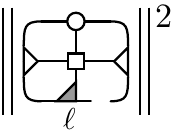}} 
 , 
 \\ 
\variance^{(n\ge2)\perp} & = \sum_{\ell=1}^{\scripteLL-(n-1)}
  \raisebox{-5.3mm}{
 \includegraphics[width=0.333\linewidth]{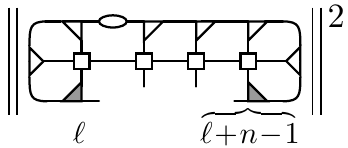}} . 
\end{flalign}
\end{subequations}
The second equality in Eq.~\eqref{eq:Variance1SitePerpDiagrams} follows from Eq.~\eqref{eq:orthogonality}.
To compute these expressions in practice, the $\D$ projectors are
expressed through $\K$ projectors using \Eq{eq:completeness-1}, e.g.\ 
\begin{align}
	 \variance^{1\perp} & = \sum_{\ell=1}^{\scripteLL}
 \raisebox{-7.5mm}{
 \includegraphics[width=0.333\linewidth]{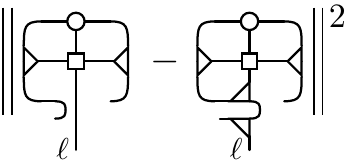}} .
\end{align}

If the Hamiltonian contains only local and nearest-neighbor terms, all contributions with 
$n\!>\!2$ are zero~\cite{Hubig2018}, i.e.\ $\variance = \variance^{\mtwosite}$. However, it has been argued in Ref.~\onlinecite{Hubig2018} that 
even if long-range terms are present, $\variance^{\mtwosite}$ is a reliable error measure. Here, we confirm this for the case of
the spin-$\frac{1}{2}$ Haldane-Shastry model on a ring of length $\eLL=40$, 
with Hamiltonian
\begin{align}
\Hc_\textrm{HS} = \sum_{\ell<\ell'\leq \seLL} \frac{\pi^2{\bf{S}}_{\ell}\cdot{\bf{S}}_{\ell'} }{\eLL^2{\rm{sin}}^2\frac{\pi}{\seLL}(\ell-\ell')}.
\label{Eq:HS2}
\end{align}
Figure~\ref{fig:variance_HS} shows $\variance^{n\perp}$ for $n\in\{1,2,\dots,10\}$ and four choices of $\Dast$. In all cases, 
$\variance^{n\perp}$ is largest for $n=2$, and smaller by an order magnitude or more for $n>2$, with the decrease being stronger the larger $\Dast$. For this model, therefore, $\Delta_E^{2\perp}$ by itself
suffices to reliably estimate the energy error.

\section{$n$-site excitations}
\label{sec:n-siteExcitations}

The \nsite\ projectors can be used as an Ansatz to compute low energy excitations. This so-called excitation Ansatz has been very successful in infinite systems~\cite{Haegeman2012,Haegeman2013a,Haegeman2013,Vanderstraeten2019,Tu2021} and lately also shown to be reliable on finite lattices~\cite{VanDamme2021}. Using our diagrammatic notation, we generalize the 
\onesite\  Ansatz for finite systems used in Ref.~\onlinecite{VanDamme2021} to $n$ sites, similar to the \nsite\ Ansatz for infinite systems~\cite{Haegeman2013a,Haegeman2013}. 
We seek an \nsite\ excitation Ansatz satisfying
the condition $\Pnsite  \ket{\Psi^{\mnsite}_{\mr{ex}}} = \ket{\Psi^{\mnsite}_{\mr{ex}}}$. Let us choose 
$\ellp\!=\!\eLL\!-\!n\!+\!1$ in
Eq.~\eqref{eq:PnsiteDefinition},  
such that $\Pnsite = \sum_{\ell=1}^{\scripteLL-n}  \Pnsite_{\ell < } + \Pnsite_{\ellp}$. Then, the following  Ansatz has the desired property:
\begin{align}
\nonumber
\ket{\Psi^{\mnsite}_{\mr{ex}}} 
&=
\sum_{\ell=1}^{\scripteLL-n}  \!\!\!
\raisebox{-0.5\height}{
 \includegraphics[width=0.322\linewidth]{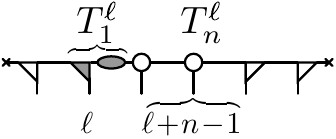}} 
+ \!\! \raisebox{-0.5\height}{
 \includegraphics[width=0.34\linewidth]{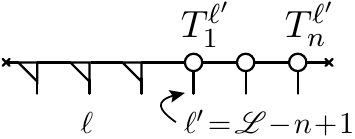}} 
\\ 
\label{eq:nsite_exAnsatz}
&=
\sum_{\ell=1}^{\scripteLL-n+1}  \!\!\!
\raisebox{-0.5\height}{
 \includegraphics[width=0.322\linewidth]{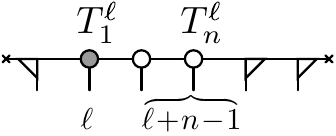}} 
 \, . 
\end{align}
Here, 
$T^{\ell}_{i>1}$ (\CircleWhiteC) are generic tensors of rank 3 and 
\begin{align}
\label{eq:TL1_def}
T^\ell_1  = \raisebox{-1mm}{\CircleGreyC} = \begin{cases}
\TriangleEllipseGrey   & \ell < \ell' \, , 
\\
\CircleWhiteC  &  \ell = \ell' \, ,  
\end{cases} \qquad  \ell' \!=\! \eLL\!-\! n \! + \! 1 \, . 
\end{align}
The two forms of $T^\ell_1$
reflect the presence or absence of a $\D$ projection 
associated with $\Pnsite_{\ell < }$ or $\Pnsite_{\ellp}$, 
respectively.
It seems that $\ket{\Psi^{\mnsite}_{\mr{ex}}}$ cannot be efficiently computed, since it involves a sum over
$\eLL-n+1$ (i.e.\ many!) terms, and performing MPS sums explicitly 
leads to increased bond dimensions.
However, that can be avoided here.
The isometries $A_{\ell}$ (\TriangleWhiteA) and $B_{\ell}$ (\TriangleWhiteB) flanking the modified sites reappear in every summand and only need to be saved once; hence only the tensors $T^{\ell}_{i}$ need to be saved. In the case of $n=1$ for example, we have to save $\eLL$ tensors of dimensions $D\times d\times D$, i.e.\ the same memory requirement as for an MPS with  bond dimension $D$.

Moreover, Eq.~\eqref{eq:TL1_def} ensures that all summands are by construction  mutually orthogonal, facilitating the computation of
overlaps.
Consider $\ket{\Psi^{\mnsite}_{\mr{ex}}}$ and $\ket{\Psi^{\prime\mnsite}_{\mr{ex}}}$, characterized by $T^{\ell}_{i}$ and $T^{\prime\ell}_{i}$, respectively. 
Due to \Eq{eq:TL1_def}, their overlap involves only $\eLL\!-n\!+\!1$ terms
(not that number squared), namely
\begin{align}
\label{eq:Ex_overlap}
\langle \Psi^{\prime\mnsite}_{\mr{ex}} | \Psi^{\mnsite}_{\mr{ex}} \rangle
= \sum_{\ell=1}^{\scripteLL-n+1}
\raisebox{-0.45\height}{
 \includegraphics[width=0.167\linewidth]{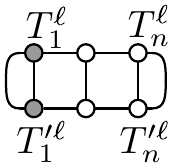}} \, ,
\end{align}
while the computation of sums or differences can be done on the level of the $T^{\ell}_{i}$, i.e.
\begin{align}
\label{eq:PsiNsiteAddition}
\ket{\Psi^{\mnsite}_{\mr{ex}}} + a\ket{\Psi^{\prime\mnsite}_{\mr{ex}}}
\to \forall \ell: 
\raisebox{-0.5\height}{
 \includegraphics[width=0.433\linewidth]{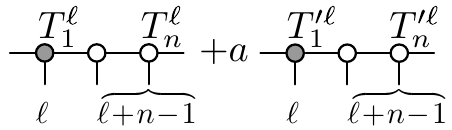}}
 \, .
\end{align}
If $\prod_{i=1}^{n} T^{\ell}_i$ and $\prod_{i=1}^{n} T^{\prime\ell}_i$ are represented as MPSs, \Eq{eq:PsiNsiteAddition} 
in effect involves a sum of two \nsite\  MPS;
this is manageable if $n$ is not too large. In the case $n=1$, there is only $T^{\ell}_1$ and $T^{\prime\ell}_1$, i.e.\ in this case, no MPS sums are required. 

A further benefit of Eq.~\eqref{eq:TL1_def} is that it serves
to fix the MPS gauge degree of freedom on the site hosting $T^1_\ell$,
improving numerical stability.

To determine the tensors $T_i^\ell$ for $\ket{\Psi^{\mnsite}_{\mr{ex}}}$ explicitly,  one 
projects the Hamiltonian onto the space $\doubleVnsite$ 
and solves for low-energy states of 
\begin{align}
\label{eq:Exn_eigenvalueEq}
\Pc^{\mnsite}H\Pc^{\mnsite}\ket{\Psi^{\mnsite}_{\mr{ex}}} = E^{\mnsite}_{\mr{ex}} \ket{\Psi^{\mnsite}_{\mr{ex}}} 
\end{align}
that are orthogonal to the ground state. This can be done using some iterative eigensolver like the Lanczos method, initialized by some appropriate initial wavefunction. Explicit orthogonalization w.r.t.\ to the ground state is required, since our Ansatz space $\Pnsite$ contains the ground state, whose kept and discarded spaces span the image of $\Pnsite$. 

To run an iterative eigensolver, a scheme is needed
for efficiently applying the projected Hamiltonian $\Pc^{\mnsite}H\Pc^{\mnsite}$ to the state $\ket{\Psi^{\mnsite}_{\mr{ex}}}$.
The resulting
state, say $\ket{\overline{\Psi}{}^{\mnsite}_{\mr{ex}}}
= \Pc^{\mnsite}H\Pc^{\mnsite}\ket{\Psi^{\mnsite}_{\mr{ex}}}$,
will again be of the form \eqref{eq:nsite_exAnsatz},
but described by tensors $\overline{T}{}^{\ell}_{\! i}$. 
To find these, we compute the tensors
\begin{align}
\label{eq:Exn_applyH_concept}
\raisebox{-0.23\height}{
 \includegraphics[width=0.167\linewidth]{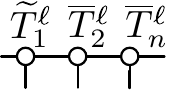}}
&= \!\!\!\sum_{\ell'=1}^{\scripteLL-n+1} \!
\raisebox{-0.38\height}{
 \includegraphics[width=0.493\linewidth]{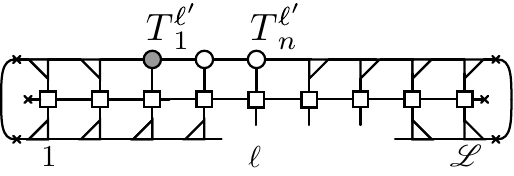}} \, , 
\end{align}
and project $\widetilde{T}{}^{\ell}_{\!1}$ to the discarded space to obtain $\overline{T}{}^{\ell}_{\!1}$,
\begin{align}
\label{eq:TL1_ortho}
\raisebox{-0.35\height}{
 \includegraphics[width=0.533\linewidth]{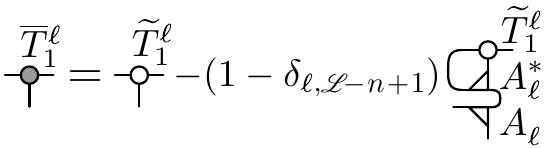}}
 \, ,
\end{align}
such that Eq.~\eqref{eq:TL1_def} is fulfilled.
To evaluate \Eq{eq:Exn_applyH_concept}, we
split the sum $\sum_\ellp$ into terms with $\ell'< \ell$ and
$\ell'\ge \ell$, and express these as follows:
\begin{align}
\label{eq:Exn_applyH_concept_result}
\raisebox{-0.23\height}{
 \includegraphics[width=0.167\linewidth]{Eq/ExnTTT}}
&= \sum_{m=1}^{n} 
\! \raisebox{-0.4\height}{
 \includegraphics[width=0.32\linewidth]{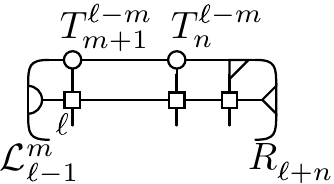}}
\hspace{-0.25cm}  
\\ \nonumber
& \; +
\sum_{m=0}^{n} 
\! \raisebox{-0.4\height}{
 \includegraphics[width=0.32\linewidth]{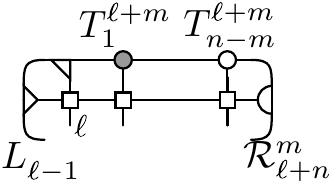}}
\hspace{-0.25cm}  
\, .
\end{align}
Next to the left and right environments 
$L_\ell$ and $R_\ell$ defined in \Eq{eq:L-R-environments}, 
these expressions contain another set of environments, denoted by $\mc{L}^{m}_{\ell}$ and $\mc{R}^{m}_{\ell}$, each involving those $m$ of the 
$T^\ellp_i$ tensors in \Eq{eq:Exn_applyH_concept} that do not face open physical legs. 
For $m=0$,  $m\! \in\! \{1,\dots,n\!-\!1\}$ or
$m=n$, they are defined by the left equalities
below; the right equalities show how for 
each $m$,  $\Lc^{m}_{\ell+1}$ and $\Rc^{m}_{\ell-1}$ can be computed recursively from $\Lc^{m}_{\ell}$ and $\Rc^{m}_{\ell}$, initialized with 
$\Lc^0_0 = 1$, $\Lc^{m>0}_0 = 0$, 
$\Rc^0_{\scripteLL+1} = 1$, $\Rc^{m>0}_{\scripteLL+1} = 0$:
\begin{subequations}
\label{eq:RLc_def}
\begin{align}
 \mc{L}^{0}_{\ell}
\,=\,
\!\raisebox{-0.36\height}{
 \includegraphics[width=0.0222\linewidth]{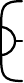}} 
 & \,= \,
 \!\!\!\raisebox{-0.55\height}{
 \includegraphics[width=0.04\linewidth]{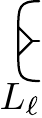}} 
 \,= \,
\hspace{-0.3cm}
 \raisebox{-0.58\height}{
 \includegraphics[width=0.13\linewidth]{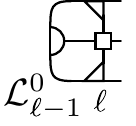}}
\hspace{0.05cm}
 \, , 
 \\ \nonumber
  \mc{L}^m_{\ell}
  \,=\,
\!\raisebox{-0.36\height}{
 \includegraphics[width=0.0222\linewidth]{Eq/Lc_ml}}
 &\,=\,
 \hspace{-0.3cm}
 \raisebox{-0.4\height}{
 \includegraphics[width=0.367\linewidth]{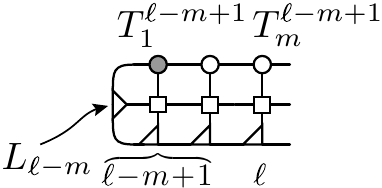}}
 \hspace{-0.7cm} 
\, = \,
\!\! \hspace{-0.1cm}
 \raisebox{-0.45\height}{
 \includegraphics[width=0.19\linewidth]{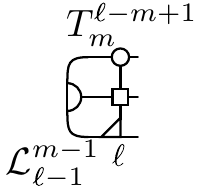}}
 \, , 
 \\ \nonumber
 \mc{L}^n_{\ell}
 \,=\,
\!\raisebox{-0.36\height}{
 \includegraphics[width=0.0222\linewidth]{Eq/Lc_ml}}
 &\,= 
\!
 \sum_{\ell'=1}^{\ell-n+1} \!\!\! 
 \!\raisebox{-0.37\height}{
 \includegraphics[width=0.27\linewidth]{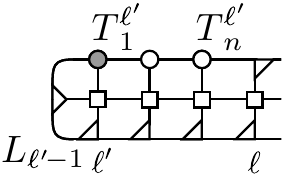}}
\,\,=
\hspace{-0.4cm}
 \raisebox{-0.43\height}{
 \includegraphics[width=0.18\linewidth]{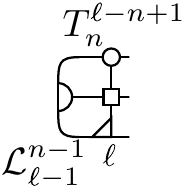}}
 \hspace{-0.3cm}
+
 \hspace{-0.3cm}
 \raisebox{-0.43\height}{
 \includegraphics[width=0.13\linewidth]{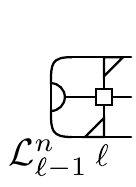}}
 \hspace{-0cm} \; , 
 \\
  \mc{R}^{0}_{\ell}
\,=\,
\!\raisebox{-0.36\height}{
 \includegraphics[width=0.0222\linewidth]{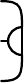}}  
 &\,= \,
 \!\!\raisebox{-0.55\height}{
 \includegraphics[width=0.04\linewidth]{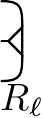}} 
 \!\!
 \,=\,
 \!\!\!
 \raisebox{-0.58\height}{
 \includegraphics[width=0.13\linewidth]{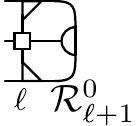}}
\hspace{0.05cm} \!, 
\\ \nonumber
  \mc{R}^m_{\ell}
 \, =\,
\!\raisebox{-0.36\height}{
 \includegraphics[width=0.0222\linewidth]{Eq/Rc_ml}}
 &\,=\,
\!\!\!\!
 \raisebox{-0.4\height}{
 \includegraphics[width=0.333\linewidth]{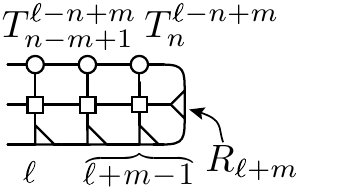}}
\hspace{-0.7cm}
\,=
 \raisebox{-0.45\height}{
 \includegraphics[width=0.16\linewidth]{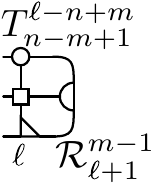}}
\, , 
 \\\nonumber
 \mc{R}^n_{\ell}
 \,=\,
\!\raisebox{-0.36\height}{
 \includegraphics[width=0.0222\linewidth]{Eq/Rc_ml}}
 &\,=  \,
 \hspace{-0.2cm}
 \sum_{\ell'=\ell}^{\scripteLL-n+1} \hspace{-0.1cm}
\raisebox{-0.39\height}{
 \includegraphics[width=0.28\linewidth]{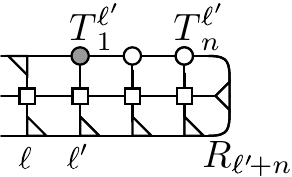}}
 \hspace{-0.2cm}
\,= 
\hspace{-0.0cm}
 \raisebox{-0.41\height}{
 \includegraphics[width=0.14\linewidth]{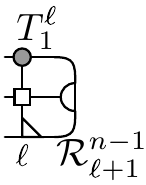}}
 \hspace{-0.3cm} 
 + \hspace{-0cm}
 \raisebox{-0.41\height}{
 \includegraphics[width=0.14\linewidth]{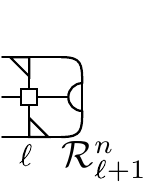}}
\hspace{-0.3cm}  
\, .
\hspace{-10cm}& &
\end{align}
 \end{subequations}
 
 The solution of \Eq{eq:Exn_eigenvalueEq} using an iterative eigensolver has costs scaling with $\mc{O}(D^3 d^n w)$, the same as \nsite\ DMRG. However, because the Ansatz \Eq{eq:nsite_exAnsatz} is built from a sum over $\eLL-n+1$ MPSs, states can be captured which would need significantly larger bond dimensions if represented in standard fashion 
 as an MPS. 
 Because there are $n$ summands in \Eq{eq:nsite_exAnsatz} which differ from the ground state at site $\ell$ (with corresponding tensors $T_1^{\ell},\dots,T_n^{\ell-n+1}$ at site $\ell$), 
 an MPS representation would need  bond dimension 
 $D(1+n)$,
 assuming $A_{\ell}$, $B_{\ell}$ and $T^{\ell}_{i}$ are 
 tensors of dimension $D\times d\times D$. Optimizing such an MPS with \nsite\ DMRG comes with $\mathcal{O}(D^3 (n+1)^3 d^n w)$ costs,
 larger by $(n+1)^3$ than the costs for optimizing the Ansatz \Eq{eq:nsite_exAnsatz}. Of course, the latter Ansatz is much more restrictive than a generic MPS of bond dimension $D(1+n)$. However, that should not be a limitation if the physics of interest involves single- or few-particle
 excitations, as is the case, e.g.,\ when computing correlations
 functions of single- or few-particle operators.
 %
 
 %
 
\begin{figure}[t!]
 \includegraphics[width=\linewidth]{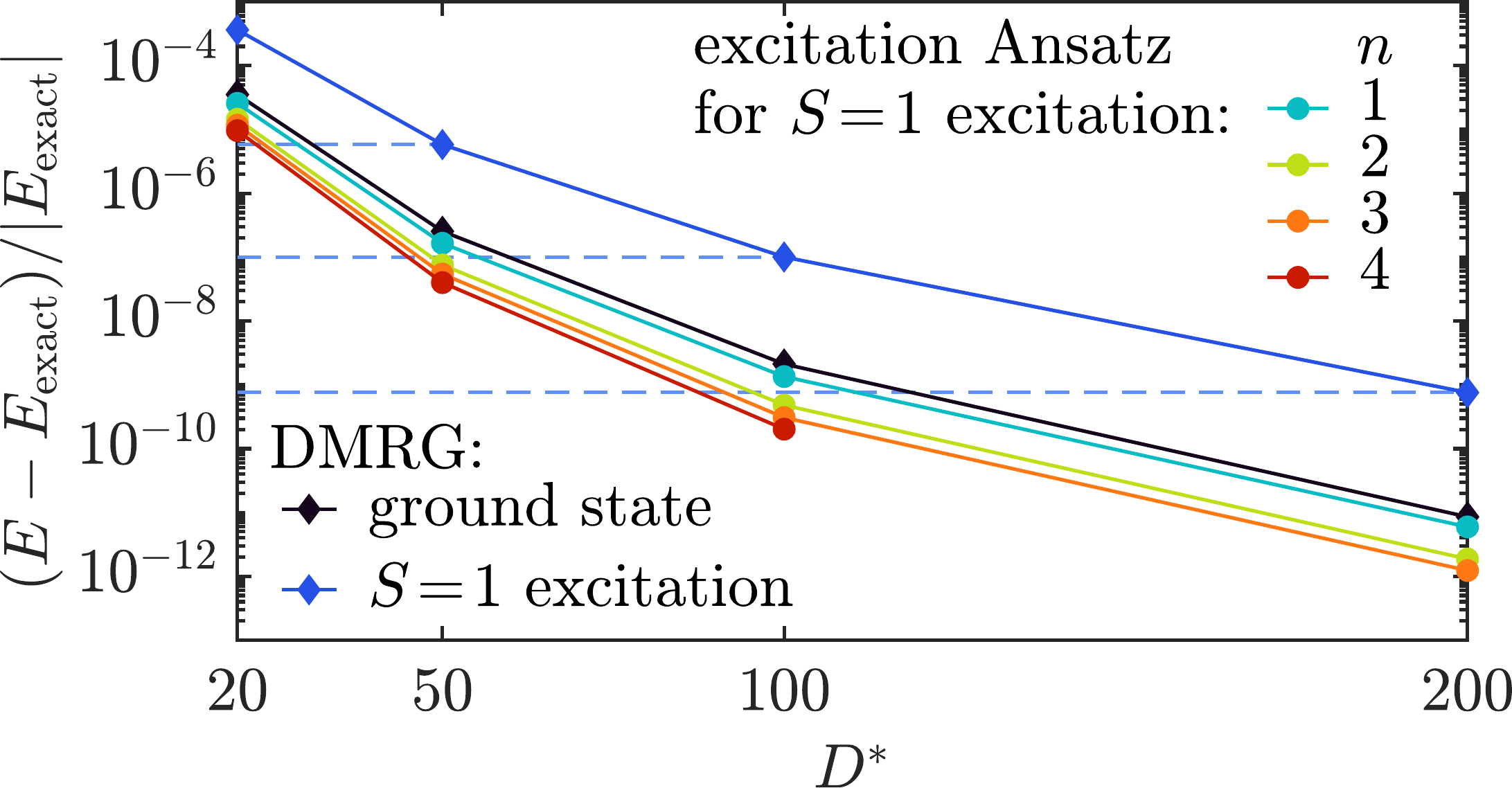} 
  \vspace{-7mm}
 \caption{
Relative error in energy of the lowest-lying $S\!=\!1$ excited state of the
Haldane-Shastry model, computed using the $n$-site excitation Ansatz (circles),
or using DMRG (blue diamonds). Black diamonds show
DMRG results for the $S\!=\!0$ ground state. 
The dashed blue lines are guides to the eye.
 \label{fig:excitation_HS}} 
 \end{figure}

 We test the \nsite\ excitation Ansatz on a Haldane-Shastry model on a ring of length $\eLL\!=\!40$ (see \Eq{Eq:HS2} for the Hamiltonian), for which we seek to compute the lowest energy excitation with total spin $S\!=\!1$ above the total spin $S\!=\!0$ ground state. For comparison, we have also computed this state 
 by performing a DMRG ground state search in the $S=1$ sector.

 Fig.~\ref{fig:excitation_HS} shows the corresponding relative errors in energy versus the bond dimension $D^\ast$. As reference values, we use the exact energies 
 $E_{\mr{exact}}^{S=0} = -\pi^2(\eLL+5/\eLL)/24$ and $E_{\mr{exact}}^{S=1}= -\pi^2(\eLL-7/\eLL)/24$ for the ground state and excited state~\cite{Yamamoto2000,Yamamoto2000a,Wu2020}, respectively. Remarkably, we find that 
 for the same $\Dast$, the $n\!=\!1$ site excitation Ansatz yields an
 $S=1$ excitation energy that is more accurate than that obtained
 from DMRG by one to two orders of magnitude, even though the computational cost of both approaches at the same $D^\ast$ is comparable. In fact, the relative error 
 obtained by the excitation Ansatz for the
 $S=1$ state is comparable to (even slightly lower than) that obtained by DMRG for the  $S=0$ ground state. 

The reason for the high accuracy of the excitation Ansatz is that the first excited state of the Haldane-Shastry model is 
essentially
a superposition of local spin excitations, i.e.\ it fits Ansatz~\eqref{eq:nsite_exAnsatz}. 
The excitation Ansatz avoids representing this superposition as a single MPS, which would require about twice the bond dimension.
Instead, it exploits the fact that each 
local excitation differs from the ground state only locally. This leads to a more economic Ansatz compared to DMRG, 
which needs about twice
the bond dimension. This can also be seen in Fig.~\ref{fig:excitation_HS}, where the relative error in energy of the \onesite\ excitation Ansatz at some $D^\ast$ almost coincides with the corresponding error of DMRG at $2 D^\ast$. 
The latter error is slightly smaller than the former, because the $2 D^\ast$ MPS Ansatz used by DMRG is less restrictive than the $D^\ast$ excitation Ansatz, though this improvement is rather marginal.
The capability of the excitation Ansatz can be further improved by considering $n>1$,
leading to a reduction of the relative error in energy compared to $n\!=\!1$, see  Fig.~\ref{fig:excitation_HS}. This reduction is rather small and further improvements seem to become ever smaller for ever larger $n$. However, 
with increasing $n$ the costs for this  Ansatz increase exponentially, 
as $\sim d^n$. Therefore, including information beyond $n\!=\!1$ by brute force, i.e.\ by just going to $n\!>\!1$, is not advisable.  Nevertheless, we believe that 
valuable improvements of the Ansatz may be achievable, while circumventing the exponential $d^n$ scaling,  by including 
only those parts of the $n> 1$ sectors that contribute to the
excited state with significant weight. It should be possible to identify these parts by generalizing the strategy proposed in our recent work on controlled bond expansion in both DMRG ground state search~\cite{Gleis2022} and TDVP time evolution~\cite{Li2022}. We leave this as a topic for future study.

More generally,  we believe that the diagrammatics for the $n$-site excitation Ansatz and the projector formalism developed in this work will provide a solid foundation to construct systematic improvements to the 1-site excitation Ansatz without a significant increase in computational costs.

We conclude this section by noting that the above construction will not be able to find states that differ from a given ground state on an extensive number of sites. 
In particular, if the ground state sector has a degeneracy, e.g. due to symmetry breaking or topological order, the excitation Ansatz on top of one of the ground states is not expected to reliably find the other ground states. 

Further, while the excitation Ansatz Eq.~\eqref{eq:nsite_exAnsatz} can in principle be used for excitations at any energy, it is expected to perform less reliable the higher the energy of the excitation. Examples, where the Ansatz Eq.~\eqref{eq:nsite_exAnsatz} should have problems, are excitations of multiple independent particles (i.e. the particles may be located far apart from each other) or excited states with a volume-law entanglement entropy.

\section{Summmary and Outlook}
\label{sec:Outlook}

We have developed a projector formalism for kept and discarded spaces of MPS, together with a convenient diagrammatic notation. We use it  to derive explicit expressions for global $n$-site projectors $\Pnsite$ and irreducible $n$-site projectors $\Pc^{n\perp}$. We then use our results to derive explicit formulas for the $n$-site variance and evaluate it for the Haldane-Shastry model, showing that indeed the 2-site contribution is the most dominant one. Further, we derive explicit diagrammatic formulas to perform excited state computations based on the $n$-site excitation Ansatz for finite, non-translation invariant MPS. 

The $\K$,$\D$ projector formalism and diagrammatic notation developed here proved very convenient for the applications considered in this work. 
More generally, we expect them to provide a convenient tool for the development of new MPS algorithms that explicitly or implicitly utilize the properties of discarded spaces. The information contained in these is a \textit{resource},
useful for describing changes or variations of a given MPS, and 
for algorithms exploiting this resource, the $\K$,$\D$ projector formalism facilitates book-keeping thereof. Indeed, we have developed the formalism presented here while working out a controlled bond expansion algorithm to perform both DMRG ground-state searches~\cite{Gleis2022} and time evolutions using the time-dependent variational principle~\cite{Li2022} with 2-site accuracy at 1-site computational cost. Morever, our formalism provides the tools needed to efficiently implement the perspectives outlined in Refs.~\onlinecite{Haegeman2013a,Vanderstraeten2019}
for post-MPS applications, that build on a given MPS to compute low energy excitation spectra.

As a final remark, we note that though we focused on MPSs in this work, our formalism should be generalizable to any tensor network for which canonical forms are available, such as tensor networks without loops.

\section*{Acknowledgements}

We thank Andreas Weichselbaum for stimulating discussions,
and Seung-Sup Lee, Juan Espinoza, Matan Lotem, Jeongmin Shim and Andreas Weichselbaum for helpful comments on the manuscript. Our numerical simulations employed the QSpace tensor library \cite{Weichselbaum2012,Weichselbaum2020}. This research was funded in part by the Deutsche Forschungsgemeinschaft under Germany's Excellence Strategy EXC-2111 (Project No.\ 390814868),
and is part of the  Munich Quantum Valley, supported by the Bavarian state government with funds from the Hightech Agenda Bayern Plus.

\bibliography{KDprojectors}

\end{document}